\documentclass[journal]{IEEEtran}
\usepackage[T1]{fontenc}
\usepackage{cite}
\usepackage{amssymb}
\usepackage{amsfonts}
\usepackage{fancyhdr}  %
\usepackage{extarrows}
\usepackage{algorithm}
\usepackage{multirow,tabularx}
\usepackage{mathtools}
\usepackage{xcolor}
\usepackage[english]{babel}
\usepackage{caption}
\usepackage{bm}
\usepackage{algorithmic}
\usepackage{amsmath}
\usepackage{subfigure}
\usepackage{makecell}
\usepackage{colortbl}

\ifCLASSINFOpdf
  \usepackage[pdftex]{graphicx}
  \graphicspath{{../pdf/}{../jpeg/}}
  \DeclareGraphicsExtensions{.pdf,.jpeg,.png}
\else
  \usepackage[dvips]{graphicx}
  \graphicspath{{../eps/}}
  \DeclareGraphicsExtensions{.eps}
\fi
\begin{document}
\title{Joint Multi-User Communication and Sensing Exploiting Both Signal and Environment Sparsity}

\author{Xin~Tong,~\IEEEmembership{Student Member,~IEEE},
        Zhaoyang~Zhang,~\IEEEmembership{Senior Member,~IEEE},
        Jue~Wang,~\IEEEmembership{Student Member,~IEEE}, 
        Chongwen~Huang,~\IEEEmembership{Member,~IEEE},
        and~M\'{e}rouane~Debbah,~\IEEEmembership{Fellow,~IEEE}
\thanks{X.~Tong, Z.~Zhang (Corresponding Author), J.~Wang and C.~Huang are with the College of Information Science and Electronic Engineering, Zhejiang University, Hangzhou 310007, China, and Zhejiang Provincial Key Lab of Information Processing, Communication and Networking (IPCAN), Hangzhou 310007, China, and the International Joint Innovation Center, Zhejiang University, Haining 314400, China (e-mails: \{tongx, ning\_ming, juew, chongwenhuang\}@zju.edu.cn).}

\thanks{M. Debbah is with the Technology Innovation Institute, 9639 Masdar City, Abu Dhabi, United Arab Emirates (email: merouane.debbah@tii.ae) and also with CentraleSupelec, University Paris-Saclay, 91192 Gif-sur-Yvette, France.}

}

\maketitle
\begin{abstract}
  As a potential technology feature for 6G wireless networks, the idea of sensing-communication integration requires the system not only to complete reliable multi-user communication but also to achieve accurate environment sensing. In this paper, we consider such a joint communication and sensing (JCAS) scenario, in which multiple users use the sparse code multiple access (SCMA) scheme to communicate with the wireless access point (AP). Part of the user signals are scattered by the environment object and reflected by an intelligent reflective surface (IRS) before they arrive at the AP. We exploit the sparsity of both the structured user signals and the unstructured environment and propose an iterative and incremental joint multi-user communication and environment sensing scheme, in which the two processes, i.e., multi-user information detection and environment object detection, interweave with each other thanks to their intrinsic mutual dependence. The proposed algorithm is sliding-window based and also graph based, which can keep on sensing the environment as long as there are illuminating user signals. The trade-off relationship between the key system parameters is analyzed, and the simulation result validates the convergence and effectiveness of the proposed algorithm.
\end{abstract}

\begin{IEEEkeywords}
  JCAS, sensing-communication integration, compressed sensing, SCMA, environment sensing
\end{IEEEkeywords}

\IEEEpeerreviewmaketitle

\section{Introduction}
\subsection{Motivation}
\IEEEPARstart{T}{he} emerging of many dazzling innovative technologies such as ultra-massive MIMO, large-scale intelligent reflective surfaces (IRS) and wireless artificial intelligence (AI) \cite{Saad,Yang}, etc., boosts the rapid development of broadband wireless communications. On the other hand, in the foreseeable future, many revolutionary and challenging application scenarios such as smart city \cite{Zanella}, autonomous driving \cite{Toutouh} and unmanned aerial vehicle (UAV) positioning \cite{Bor}, etc, may call for not only broadband connections but also accurate environment information which include but are not limited to, the locations, shapes, status and electromagnetic characteristic of the stationary or moving objects and/or the background scatterers, within that environment.

Such kind of environment sensing has long been accomplished by traditional Radar technology which has been regarded as a related but separate field from communication. However, the channel state information (CSI) obtained during the communication process usually contains certain knowledge of the environment \cite{Sen, Rao}, and likewise, the environment sensing result also helps improve the accuracy of channel estimation and enhance the performance of communication \cite{Jiaoicc,Jiaotwc}. As the wireless network operates in higher frequency with wider bandwidth and deploys denser base stations with a larger amount of antennas, its longly-neglected sensing capability inherited from the intrinsic nature of electromagnetic wave propagation becomes even more explorable. The signal processing principles employed in these two fields also tend to converge. As visioned in \cite{Wild}, this may give rise to a promising new technology of JCAS or sensing-communication integration.

One major challenge in JCAS lies in the potentially large amount of unknown variables brought by the environment, even many more than those contained in the statistical channel models, which may make the problem rank-deficient and eventually insolvable. As a result, certain sparsity should be exploited. Fortunately, the targeted environment itself usually does possess certain sparsity, which in turn makes the communication channel sparse. For example, in a cellular communication network, buildings are sparsely located within the wireless network coverage, and in the indoor scenario, furniture and other items are sparsely distributed in the entire room. 

As a matter of fact, either in conventional communication or in conventional sensing, sparsity has been fully exploited to achieve better and lower-complexity solutions under the compressed sensing (CS) framework. As in wireless communications, the CS-based channel estimation approaches exploiting the channel sparsity usually exhibit superiority in signaling overhead and computational complexity \cite{Rao} compared to the conventional channel estimation approaches \textit{et al.}\cite{Sen}. Likewise, in the broad area of Radar sensing or computational imaging, the utilization of the intrinsic sparsity of objects or scatterers within an environment is key to their effective detection. Such problems are often modeled as sparse signal recovery problems based on pixel division according to the CS theory \cite{Donoho, Candes}, and solved by the widely used methods like Sparse Bayesian Learning (SBL) \cite{Zhang}, orthogonal matching pursuit (OMP) \cite{Cai}, and Generalized Approximate Message Passing (GAMP) \cite{Rangan}, and so on. This has been showcased by recent works in \cite{taoy, yaojj}, where innovative microwave computational imaging methods with the aid of intelligent reflecting surfaces (IRS) are proposed based on a fast block sparse Bayesian learning (BSBL) algorithm. However, how to explore and exploit the sparsity in the JCAS scenario still lacks adequate study.

\subsection{Related Works}
So far, there have been different sorts of attempts to implement joint environment sensing and communication. 

To list a few, in the Radar-Communication Coexistence (RCC) sort of approaches \cite{wang2008, Saruthirathanaworakun}, effective interference cancellation and management mechanism are designed to achieve flexible coexistence between the radar and the communication systems. In contrast to the RCC system, the second sort of approach, i.e., Dual-Functional Radar-Communication (DFRC), aim to achieve an integration of radar and communication through sharing a common hardware platform, with improved sensing and communication performance through collaborative operation \cite{Paul, Blunt}. In such systems, separate sensing and communication operations with shared radio resources have also been extensively studied. For instance, a multi-beam scheme is proposed in \cite{ZhangAndrew}, which uses an analog array to generate multiple beams for simultaneous communication and radar scanning. 

The third sort of works mainly concentrates on the purpose of environment sensing by pure usage of the conventional communication signals or by proper joint design of the radio signals, which can be found in \cite{AndrewEnabling} and the references therein. As in one of such works on environment sensing with the aid of real deployed communication systems, \cite{Tan}  identified the behavior of a human body by extracting the Doppler frequency shift from the CSI conveyed by the WIFI communication signals in the indoor scenario. In \cite{Daniels}, the authors used the orthogonal frequency division multiplexing communication waveform as a radar signal to achieve joint communication and environment sensing between vehicles. As an illustration of joint sensing and communication signaling, the authors in \cite{ChenPerformance} designed a cooperative sensing unmanned aerial vehicle network (CSUN) with joint sensing and communication beams based on a common transceiver device. Considering the non-ideal factors of the channel, \cite{Shahi} analyzed the communication channel capacity under the joint effect of Gaussian random noise and non-Gaussian radar sensing interference. In \cite{Ahmadipour}, the authors theoretically analyzed the performance of the JCAS system under the condition of the memoryless broadcast channel. The related systematic design rules and methodologies for signaling and processing in such a JCAS system have raised increasing research effort recently, resulting in some interesting JCAS implementations based on machine learning \cite{Aoudia}, joint data sensing and fusion \cite{Schmitt}, and time-frequency-space full dimension utilization \cite{Gaudio}, etc. 

\subsection{Main Ideas and Contributions}

In this paper, we exploit the sparsity of both the structured multi-user signals and the unstructured environment to design a low-complexity joint multi-user communication and environment sensing scheme based on microwave computational imaging.
Different from the above-mentioned application scenarios, we aim at designing a system with integrated sensing and communication capability based on existing wireless communication systems, which is capable of accomplishing the environment sensing (imaging) using the multi-user transmission signals and in turn, assisting the multi-user information detection with the channel information derived from the sensing results. To the best of our knowledge, there still lacks sufficient study on the design of such a JCAS system in the literature.

In order to achieve these goals, we employ the Sparse Code Multiple Access (SCMA) protocol \cite{Nikopour,Taherzadeh} for multi-user uplink access, and employ an IRS \cite{Garcia} to assist signal propagation and collection. SCMA is an elegant code-domain non-orthogonal multiple access (C-NOMA) method, which has extracted extensive research effort due to its superior performance and low detection complexity. In the SCMA scheme, the codebook for users to send data is sparse, and each user occupies a few but not all subcarriers. The sparsity of the user codebook effectively enhances the decoding performance of the received data. IRS is a promising technology to manipulate the electromagnetic environment with low-cost passive reflective elements by adjusting the phase of incident signals, which has been extensively used in wireless communications \cite{Huang, Garcia, Wu, Chenw}.  Based on the idea of computational imaging, IRS can cause known and diverse changes to electromagnetic signals, and such changes are beneficial to environment sensing. Therefore, IRS also exhibited great potential in environment sensing as recently described in \cite{yaojj, taoy}. But in these cases, the base station is only used as an environment sensing device instead of a communication signal transmitter. It is noteworthy that, in this work, the environment sensing is just accomplished by making use of the IRS' reflection characteristics rather than by actively manipulating it. Our method not only enables environment sensing to obtain the help of IRS but also retains the ability of IRS to assist communication.

Our design is depicted as follows. First, in the multiple access part, the user uses the SCMA protocol to communicate with the wireless AP. The signals are reflected by the IRS and then arrive at the AP. With the limited channel information obtained by an initial pilot sequence, we use the proposed SCMA-IRS-MPA algorithm (see Section. \ref{mainalgm} for details) to conduct multi-user detection based on the sparse codebook of the transmitted signals. Then a sliding-window-based environment sensing algorithm is proposed to accomplish the environment sensing (imaging) with the received signal and recovered users' data, again based on the CS principle. Note that the proposed multiple user detection algorithm requires the environment (channel) knowledge, and the proposed environment sensing algorithm also needs to know the decoded data, so the two processes, i.e., multiple access and environment sensing, rely on each other. Therefore, finally, we propose an iterative and incremental algorithm to jointly recover the users' data and accomplish environment sensing at the same time with significantly reduced pilot overhead.

The main contributions of this paper are summarized as follows:

\begin{itemize}
    \item We design a joint communication and environment sensing scheme, which exploits the sparsity of both the structured multi-user signals and the unstructured environment to achieve the integration of multiple access and environment sensing. 
    
    \item We develop a low-complexity iterative algorithm based on CS and generalized message passing theory to conduct the multi-user information detection and environment sensing (imaging). It is sliding-window based and runs alternately between the states of sensing with the decoded multi-user data and data decoding with the sensing results. This way, the overall system performance can be incrementally improved.   
    
    \item We analyze the decoding error and sensing accuracy performances as well as the computational complexity of the proposed algorithm, and investigate the impact of access user number on system performances, base on which, we approximate the optimal operating point. Extensive simulation results verify the convergence and effectiveness of the proposed algorithm.
        
\end{itemize}

The rest of this paper is organized as follows. Section \uppercase\expandafter{\romannumeral2} presents the environment setting and system model in the uplink communication scenario. Section \uppercase\expandafter{\romannumeral3} proposes the multiple access method based on the SCMA scheme. Section \uppercase\expandafter{\romannumeral4} proposes the environment sensing method base on the CS theory. Section \uppercase\expandafter{\romannumeral5} proposes the iterative and incremental algorithm based on low-density pilots jointly recovers the users' data and achieves environment sensing. In section \uppercase\expandafter{\romannumeral6}, we discuss the trade-off relationship between the number of access users and system performance. Finally, section \uppercase\expandafter{\romannumeral7} presents the numerical results, and section \uppercase\expandafter{\romannumeral8} concludes the paper.

\textit{Notation}: Fonts $a$ and $\mathbf{A}$ represent scalars and matrices, respectively.
$\mathbf{A}^{\rm{T}}$ and $\|\mathbf{A}\|_F$ denote transpose and Frobenius norm of $ \mathbf{A} $, respectively.
$[\mathbf{A}](i,j)$ represents $\mathbf{A}$'s $(i,j)$-th element.
$|\cdot|$ and $[\cdot]$ denote the modulus and the catenation of the matrix, respectively.
$\odot $ represents the Hadamard product between two matrices.
Finally, notation ${\rm diag}(\mathbf{a})$ represents a diagonal matrix with the entries of $\mathbf{a}$ on its main diagonal, and $\delta(\cdot)$ is the Dirac delta function.

\begin{figure}
  \centering
  \includegraphics[width=3in]{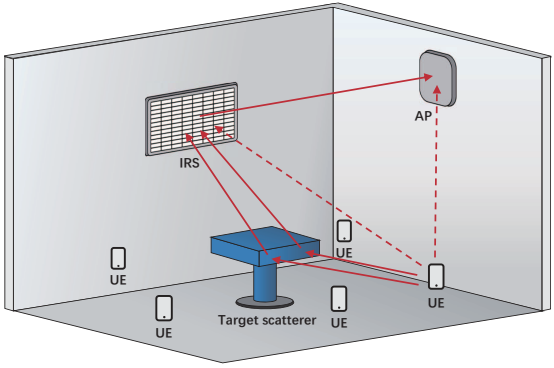}
  \caption{The uplink multi-user communication scenario.}
  \label{figsetting}
  \end{figure}

\section{Environment Setting and System Model}
As shown in Fig. \ref{figsetting}, a millimeter-wave multi-antenna AP and multiple single-antenna user equipments (UEs) are deployed in an indoor scenario. An IRS is deployed near the AP to assist the users' communication, and there are some target objects (serve as scatterers) in the scenario. We consider that in the uplink regime, i.e., multiple users simultaneously send data to the AP via a shared channel. Our goal is to accomplish the environment sensing while reliably obtaining the communication data, that is, to sense the distribution position and scattering rate of the scatterers in the environment and obtain the communication data of all the users at the same time.

\subsection{Environment Setting}
Let the number of users in the environment be $N_{\rm{u}}$ and the number of AP receiving antennas be $N_{\rm{R}}$. In the uplink communication scenario, the channel from the user to the AP has mainly composed of three parts: The first part is the line-of-sight (LOS) path from the user directly to the AP. The second part is the path from the user to the IRS and then reflected to the AP. The third part is the multipath propagation path which the user signal is scattered by the scatterers and reflected to the AP by the IRS. They are denoted as ${{\bf{H}}^{{\rm{LOS}}}} \in \mathbb{C}^{{N_{\rm{u}}} \times {N_{\rm{R}}}}$,  ${{\bf{H}}^{{\rm{IRS}}}} \in \mathbb{C}^{{N_{\rm{u}}} \times {N_{\rm{R}}}}$, and ${{\bf{H}}^{{\rm{s}}}} \in \mathbb{C}^{{N_{\rm{u}}} \times {N_{\rm{R}}}}$ respectively. Let the number of reflective elements of the IRS be $N_{\rm{I}}$, and each element can set amplitude reflection coefficient and phase shift independently, thereby controlling the relationship between the reflected signal and the incident signal. The reflection characteristic matrix of IRS is expressed as
\begin{equation}
{\bf{\Theta }} = {\rm{diag}}\left( {{\theta _1}, \cdots, {\theta _{{N_{\rm{I}}}}}} \right) \in \mathbb{C}^{{N_{\rm{I}}} \times {N_{\rm{I}}}}, \label{eq1}
\end{equation}
where ${\theta _{{n_{\rm{I}}}}} = {\rho _{{n_{\rm{I}}}}}{e^{j{\varphi _{{n_{\rm{I}}}}}}}$ represents the reflection characteristic of the $n_{\rm{I}}$ element of the IRS, ${\rho _{{n_{\rm{I}}}}} \in \left[ {0,1} \right]$ and ${\varphi _{{n_{\rm{I}}}}} \in \left[ {{\rm{0}},2\pi } \right]$ represents the amplitude reflection coefficient and phase shift of the $n_{\rm{I}}$ element respectively, and the diag function represents the construction of a diagonal matrix.

\begin{figure}
  \centering
  \includegraphics[width=2in]{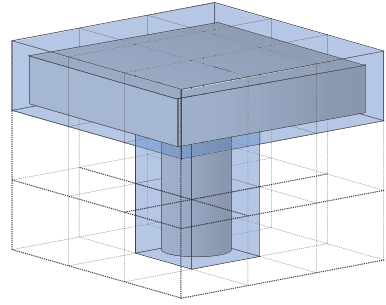}
  \caption{The discretized targeted environment object.}
  \label{pixel}
  \end{figure}

We discretize the room environment information and regard the environment information of the entire room as a point cloud. Each point in the point cloud represents the environment information of small cubes with sizes $l_{\rm{s}}$, $w_{\rm{s}}$, and $h_{\rm{s}}$ around this point. These small cubes are called pixels. Assuming that the length, width, and height of the room are $L_{\rm{s}}$, $W_{\rm{s}}$, and $H_{\rm{s}}$ respectively, the number of point clouds in the space is ${N_{\rm{s}}} = {{{L_{\rm{s}}}} \mathord{\left/
{\vphantom {{{L_{\rm{s}}}} {{l_{\rm{s}}}}}} \right.
\kern-\nulldelimiterspace} {{l_{\rm{s}}}}} \times {{{W_{\rm{s}}}} \mathord{\left/
{\vphantom {{{W_{\rm{s}}}} {{w_{\rm{s}}}}}} \right.
\kern-\nulldelimiterspace} {{w_{\rm{s}}}}} \times {{{H_{\rm{s}}}} \mathord{\left/
{\vphantom {{{H_{\rm{s}}}} {{h_{\rm{s}}}}}} \right.
\kern-\nulldelimiterspace} {{h_{\rm{s}}}}}$. The inside of each pixel may be empty, or there may be scatterers. We use a scattering coefficient ${x_{{n_{\rm{s}}}}}$ to represent the scattering coefficient of the pixel where the $n_{\rm{s}}$-th point cloud point is located. If the inside of the small cube is empty, then ${x_{{n_{\rm{s}}}}} = 0$. As shown in the Fig. \ref{pixel}, the target scatterer in Fig. \ref{figsetting} is discretized. Therefore, the environmental information of the entire room can be expressed as
\begin{equation}
{\bf{x}} = {\left[ {{x_1},{x_2}, \cdots ,{x_{{N_{\rm{s}}}}}} \right]^{\rm{T}}}. \label{eq2}
\end{equation}

\subsection{System Model}
Multiple users in the space share the same time-frequency resources. The frequency resources of communication are divided into $R$ orthogonal resource elements (OREs), and $N_u$ users send data on all OREs. Therefore, on the $r$-th ORE, the channel from the user to the AP can be expressed as
\begin{equation}
\begin{array}{l}
  {{\bf{H}}_r} = {\bf{H}}_r^{{\rm{LOS}}} + {\bf{H}}_r^{{\rm{IRS}}} + {\bf{H}}_r^{\rm{s}}\\
  \quad  = {\bf{H}}_r^{{\rm{LOS}}} + {\bf{H}}_r^{{\rm{IRS1}}}{\bf{\Theta H}}_r^{\rm{s1}}\\
  \quad  + \left[ {\begin{array}{*{20}{c}}
  {{\bf{H}}_r^{\rm{s}}\left(1\right)} & \cdots & {{\bf{H}}_r^{\rm{s}}\left(n_{\rm{u}}\right)}& \cdots &{{\bf{H}}_r^{\rm{s}}\left(N_{\rm{u}}\right)}
  \end{array}} \right]
  \end{array}, \label{eq3}
\end{equation}
where
\begin{equation}
{\bf{H}}_r^{\rm{s}}\left(n_{\rm{u}}\right) = {{\bf{x}}^{\rm{T}}}{\rm{diag}}\left( {{\bf{H}}_r^{\rm{s3}}}\left(n_{\rm{u}}\right) \right){\bf{H}}_r^{\rm{s2}}{\bf{\Theta H}}_r^{\rm{s1}} \in \mathbb{C}^{{\rm{1}} \times {N_{\rm{R}}}} \label{eq4}
\end{equation}
represents the channel coefficient from the $n_u$-th user to the AP after being scattered by the scatterers, and
${\bf{H}}_r^{{\rm{LOS}}} = {\bm{\alpha }}_r^{{\rm{LOS}}}\odot   {e^{j{\bm{\varphi }}_r^{{\rm{LOS}}}}} \in \mathbb{C}^{{{N_{\rm{u}}}} \times {N_{\rm{R}}}}$ represents the LOS channel coefficient from the user directly to the AP, with ${\bm{\alpha }}_r^{{\rm{LOS}}}$ denoting the amplitude of the channel and ${e^{j{\bm{\varphi }}_r^{{\rm{LOS}}}}}$ its phase shift. Similarly, ${\bf{H}}_r^{{\rm{IRS1}}} = {\bm{\alpha }}_r^{{\rm{IRS1}}}\odot {e^{j{\bm{\varphi }}_r^{{\rm{IRS1}}}}} \in \mathbb{C}^{{{N_{\rm{u}}}} \times {N_{\rm{I}}}}$, ${{\bf{H}}_r^{{\rm{s1}}}} = {\bm{\alpha }}_r^{\rm{s1}}\odot {e^{j{\bm{\varphi }}_r^{\rm{s1}}}} \in \mathbb{C}^{{{N_{\rm{I}}}} \times {N_{\rm{R}}}}$, ${\bf{H}}_r^{\rm{s3}}\left(n_{\rm{u}}\right) = {\bm{\alpha }}_r^{\rm{s3}}\left(n_{\rm{u}}\right)\odot {e^{j{\bm{\varphi }}_r^{{\rm{s3}}}\left(n_{\rm{u}}\right)}} \in \mathbb{C}^{{\rm{1}} \times {N_{\rm{s}}}}$, ${\bf{H}}_r^{{\rm{s2}}} = {\bm{\alpha }}_r^{{\rm{s2}}}\odot {e^{j{\bm{\varphi }}_r^{{\rm{s2}}}}} \in \mathbb{C}^{{{N_{\rm{s}}}} \times {N_{\rm{I}}}}$ represent the LOS channel coefficient from the user to the IRS, from the IRS to the AP, from the user to the spatial point cloud location, and from the spatial point cloud location to the IRS, respectively.

Let the $n_{\rm{u}}$-th user's transmitted symbols on $R$ OREs be ${{\bf{s}}_{{n_{\rm{u}}}}} \in \mathbb{C}^{{{R}} \times {1}}$, then at the $n_{\rm{R}}$-th AP receiving antenna, the received data on all OREs can be expressed as
\begin{equation}
{{\bf{y}}_{{n_{\rm{R}}}}} = \sum\limits_{{n_{\rm{u}}} = 1}^{{N_{\rm{u}}}} {{\rm{diag}}\left\{ {{{\bf{H}}\left({n_{\rm{u}}},{n_{\rm{R}}}\right)}} \right\}{{\bf{s}}_{{n_{\rm{u}}}}}}  + {\bf{w}}, \label{eq5}
\end{equation}
where ${{\bf{H}}\left({n_{\rm{u}}},{n_{\rm{R}}}\right)} = \left[ {\begin{array}{*{20}{c}}
  {{{\bf{H}}_1}\left( {{n_u},{n_R}} \right)}& \cdots &{{{\bf{H}}_R}\left( {{n_u},{n_R}} \right)}
  \end{array}} \right]$ , ${\bf{H}}_r\left(n_{\rm{u}},n_{\rm{R}}\right) $ represents the $n_{\rm{R}}$ column and $n_{\rm{u}}$ row of ${{\bf{H}}_r}$ calculated in (\ref{eq3}), and $\bf{w}$ the Gaussian white noise.

\section{The Multiple Access Scheme}
In order to recover multi-user communication data accurately, the wireless access algorithm can be used to achieve the separation and detection of multi-user transmission symbols under the premise of environmental prior information.

\subsection{Sparse Code Multiple Access}
\begin{figure*}
  \centering
  \includegraphics[width=6in]{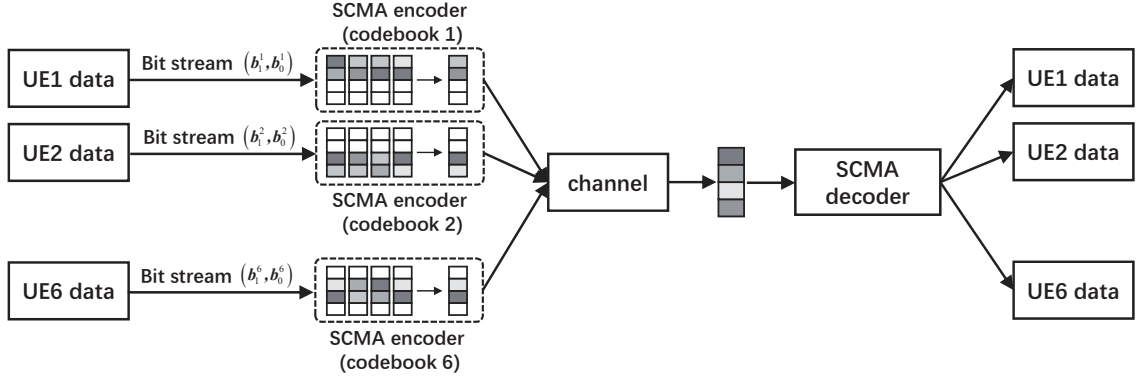}
  \caption{The uplink SCMA system ($N_{\rm{u}} = 6$, $R = 4$).}
  \label{figSCMA}
  \end{figure*}

SCMA is an efficient code-domain non-orthogonal multiple access technology. It is based on low-density spectrum spreading. That means, a single user does not completely occupy all OREs, but only a few of them, which greatly reduces the difficulty of signal decoding. In the uplink multi-user SCMA communication scenario considered in this article, the $N_{\rm{u}}$ users use the SCMA protocol to send their data to the AP and $N_{\rm{u}}$ users share $R$ OREs. Each user has a total of $M$ input possibilities, and each user accesses the channel by using a unique sparse codebook ${{\bf{C}}_{{n_{\rm{u}}}}} \in \mathbb{C}^{{{R}} \times {M}}$. Therefore, each user's codebook contains $M$ codewords. Let ${{\bf{C}}_{{n_{\rm{u}}}}\left(m\right)} \in \mathbb{C}^{{{R}} \times {1}}$ represent the $m$-th codeword of user $n_{\rm{u}}$. Then the (\ref{eq5}) can be expressed as
\begin{equation}
{{\bf{y}}_{{n_{\rm{R}}}}} = \sum\limits_{{n_{\rm{u}}} = 1}^{{N_{\rm{u}}}} {{\rm{diag}}\left\{ {{{\bf{H}}\left({{n_{\rm{u}}},{n_{\rm{R}}}}\right)}} \right\}{{\bf{C}}_{{n_{\rm{u}}}}\left(m\right)}}  + {\bf{w}}, \label{eq6}
\end{equation}
where ${{\bf{C}}_{{n_{\rm{u}}}}\left(m\right)}$ represents the sending symbols, it contains $d_{\rm{v}}$ non-zero elements, that is, each user will only transmit on the OREs represented by $d_{\rm{v}}$ non-zero elements. $N_{\rm{u}}$ users perform overload transmission on all OREs, and the number of users $d_{\rm{f}}$ transmitted on each ORE is constant. Since the codewords $\bf{C}$ is sparse, not all users' codewords will collide on a single ORE. Fig. \ref{figSCMA} shows an example of an uplink SCMA system, in which 6 users transmit on 4 OREs, thus $N_{\rm{u}}$ = 6, $R$ = 4. Each user has its own codebook, and the codebook determines the OREs occupied by the user. In Fig. \ref{figSCMA}, user 1 transmits on ORE 1 and 2, and user 2 transmits on ORE 3 and 4. The user's bitstream is mapped to the codewords by SCMA encoder after channel-coded, then transmitted to the receiver through the channel, and finally separated and decoded by the SCMA detector.

Codebook design is an important physical layer technology in the SCMA system. The sparsity of the codebook makes it possible for the SCMA receiver to use message passing algorithm (MPA) to decode. As mentioned above, the process of SCMA encoding is the process of mapping the binary bit stream to the complex domain. The codebook of each user is an ${R}\times{M}$-dimensional matrix. Therefore, the SCMA encoder can be defined as: $f:\mathbb{B}^{{{\log }_2}M} \to {\cal X}$, where. ${\cal X} \subset \mathbb{C} ^R,\left| {\cal X} \right| = M$. Let $\bf{b}$ represent the user's input bits, the corresponding codeword output can be expressed as ${\cal X} = f\left( {\bf{b}} \right)$, the codeword ${\cal X}$ is an $R$-dimensional sparse complex vector, and the vector contains $N_{\rm{c}} < R$ non-zero elements. Since SCMA encoding process combines bit-to-constellation mapping and spreading spectrum, the bit-to-constellation mapping can be expressed as: $g:\mathbb{B} ^{{{\log }_2}M} \to {\cal C},\;\;{\cal C} \subset \mathbb{C} ^{N_{\rm{c}}}$, where ${\cal C}$ represents the constellation point of the $N_{\rm{c}}$-dimensional complex constellation, so the SCMA encoder can also be expressed as : $f = {\bf{V}}g$, where ${\cal C} = f\left( {\bf{b}} \right)$ and ${\bf{V}} \in \mathbb{B} ^{R \times N}$ is a binary mapping matrix, and the mapping matrix can map $N_{\rm{c}}$-dimensional constellation points to $R$-dimensional SCMA codewords. Meanwhile, the mapping matrix of each user is different, and contains $R-N_{\rm{c}} $ all-zero rows.

Define the codebook structure of SCMA as ${\cal S}\left( {{\cal V},{\cal G};{N_{\rm{u}}},M,N_{\rm{c}},R} \right)$, where ${\cal V}: = \left[ {{{\bf{V}}_{{n_{\rm{u}}}}}} \right]_{{n_{\rm{u}}} = 1}^{{N_{\rm{u}}}}$, ${\cal G}: = \left[ {{g_{{n_{\rm{u}}}}}} \right]_{{n_{\rm{u}}} = 1}^{{N_{\rm{u}}}}$. Therefore, the SCMA codebook design problem can be expressed as
\begin{equation}
{{\cal V}^*},{{\cal G}^*} = \arg \mathop {\max }\limits_{{\cal V},{\cal G}} {\cal M}\left( {{\cal S}\left( {{\cal V},{\cal G};{N_{\rm{u}}},M,N_{\rm{c}},R} \right)} \right), \label{eq7}
\end{equation}
where ${\cal M}$ is a codebook design standard. Since there is no unified design standard at present, there are many methods for SCMA codebook design problems, such as rearranging the real and imaginary parts of the constellation points and designing codebooks based on theories such as constellation interleaving and rotation. These methods can achieve a suboptimal solution to the SCMA codebook design problem.

\subsection{SCMA-IRS-MPA Decoder}\label{mainalgm}
In the above-mentioned SCMA-IRS uplink transmission scheme, there are a total of ${M^{{N_{\rm{u}}}}}$ combinations of user codewords. The Maximum Likelihood (ML) decoder can provide the theoretically optimal symbol error rate (SER) performance by performing a traversal search on all codewords combinations. The estimated transmit codewords of all users by the ML decoder can be expressed as
\begin{equation}
\begin{array}{l}
  {{{\bf{\hat C}}}_{{\rm{ML}}}} = \arg \mathop {\min }\limits_{j \in {M^{{n_{\rm{u}}}}}} \\
  \quad \quad {\left\| {{{\bf{y}}_{{n_{\rm{R}}}}} - \sum\limits_{{n_{\rm{u}}} = 1}^{{N_{\rm{u}}}} {\left( {{\rm{diag}}\left( {{\bf{H}}\left( {{n_{\rm{u}}},{n_{\rm{R}}}} \right)} \right){{\bf{C}}_{{n_{\rm{u}}}}}\left( {{\bf{m}}\left( j \right)} \right)} \right)} } \right\|^2}
  \end{array}, \label{eq8}
\end{equation}
where ${{\bf{\hat C}}_{\rm{ML}}} = \left[ {{{{\bf{\hat c}}}_{1}}, \cdots ,{{{\bf{\hat c}}}_{{N_{\rm{u}}}}}} \right] \in \mathbb{C} ^{R \times {N_{\rm{u}}}}$, ${\bf{m}}(j)$ represents the value of the $j$-th combination among ${M^{{N_{\rm{u}}}}}$ user codewords combinations. Although the ML decoder can provide the theoretical optimal value, it uses an exhaustive method to search for the optimal solution, which is impractical in the actual implementation process. MPA decoder is an iterative decoder, which can nearly achieve the performance of an ML decoder while requiring an achievable computational complexity. And the MPA decoder obtains the corresponding user codewords by calculating the maximum joint message probability.
\begin{figure}
  \centering
  \includegraphics[width=7cm]{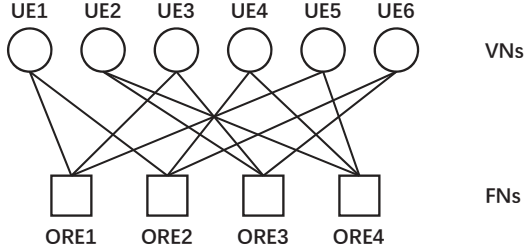}
  \caption{The MPA decoder factor graph ($N_{\rm{u}} = 6$, $R = 4$).}
  \label{figFG}
  \end{figure}

MPA is a belief propagation algorithm that uses a factor graph model to solve probabilistic reasoning problems. The proposed SCMA-IRS-MPA uses the factor graph method shown in the Fig. \ref{figFG}, where the function nodes (FNs) represent OREs, the variable nodes (VNs) represent the users, and the connection between the FN and VN represents the user transmitting data on the corresponding ORE. The MPA decoder achieves decoding by iteratively updating the message probability between FNs and VNs, and let the MPA decoder stop after ${K_{\rm{it}}}$ iterations. In order to estimate the transmission codewords in the SCMA-IRS-MPA scheme, we modified the traditional SCMA-MPA. In our method, multiple antennas at the AP perform independent and parallel decoding. For the $n_{\rm{R}}$-th receiving antenna, we use $p_{{v_u} \to {f_r}}^{\left( {{k_{\rm{it}}}} \right)}\left( {{{\bf{C}}_{{n_{\rm{u}}}}\left(m,r\right)}} \right)$ to denote the probability of transmitting a message from the $n_{\rm{u}}$-th VN to the $r$-th FN, and use $p_{{f_r} \to {v_u}}^{\left( {{k_{\rm{it}}}} \right)}\left( {{{\bf{C}}_{{n_{\rm{u}}}}\left(m,r\right)}} \right)$ to denote the probability of transmitting a message from the $r$-th FN to the $n_{\rm{u}}$-th VN. The above all represent the probability in the ${K_{\rm{it}}}$ round iteration, ${k_{\rm{it}}} = 1,2, \cdots ,{K_{\rm{it}}}$. Assuming that at the beginning, in the first iteration, all messages sent from VN to FN have the same probability,
\begin{equation}
p_{{v_u} \to {f_r}}^{\left( 0 \right)}\left( {{{\bf{C}}_{{n_{\rm{u}}}}\left(m,r\right)}} \right) = \frac{1}{M},\left( {\forall {n_{\rm{u}}},\forall r,\forall m} \right). \label{eq9}
\end{equation}

Therefore, $p_{{f_r} \to {v_u}}^{\left( {{k_{\rm{it}}} + 1} \right)}\left( {{{\bf{C}}_{{n_{\rm{u}}}}\left(m,r\right)}} \right)$ can be expressed as
\begin{equation}
\begin{array}{l}
  p_{{f_r} \to {v_u}}^{\left( {{k_{\rm{it}}} + 1} \right)}\left( {{{\bf{C}}_{{n_{\rm{u}}}}\left(m,r\right)}} \right){\rm{ = }}\\
  \quad \quad \sum\limits_{\psi \left( i \right),i \in {{\bf{\Lambda }}_r}\backslash {n_{\rm{u}}}} {\left\{ {p\left( {{\bf{y}}|\psi \left( i \right),\psi \left( u \right) = {{{\bf{C}}_{{n_{\rm{u}}}}\left(m,r\right)}}} \right)} \right.} \\
  \quad \quad \left. { \times \prod\limits_{i \in {{\bf{\Lambda }}_r}\backslash {n_{\rm{u}}}} {p_{{v_i} \to {f_r}}^{\left( {{k_{\rm{it}}}} \right)}} \left( {\psi \left( i \right)} \right)} \right\},\left( {\forall m,\forall r,{n_{\rm{u}}} \in {{\bf{\Lambda }}_r}} \right)
  \end{array}, \label{eq10}
\end{equation}
where ${{\bf{\Lambda }}_r}$ represents a set of user indexes sharing the $r$-th ORE, ${{\bf{\Lambda }}_r}\backslash {n_{\rm{u}}}$ represents ${{\bf{\Lambda }}_r}$ except for the $n_{\rm{u}}$-th user, and
\begin{equation}
\begin{array}{l}
  p\left( {{\bf{y}}|{{\bf{\Psi }}_r}} \right) = \frac{1}{{\sqrt {2\pi } \sigma }}\exp \left( { - \left| {{{\bf{y}}_r} - \sum\nolimits_{{n_{\rm{u}}} \in {{\bf{\Lambda }}_r}} {\left( {{\bf{H}}_r^{{\rm{LOS}}}\left( {{n_{\rm{u}}},{n_{\rm{R}}}} \right)} \right.} } \right.} \right.\\
   \quad \left. {{{{{\left. {\left. { + {\bf{H}}_r^{{\rm{IRS}}}\left( {{n_{\rm{u}}},{n_{\rm{R}}}} \right) + {\bf{H}}_r^{\rm{s}}\left( {{n_{\rm{u}},n_{\rm{R}}}} \right)} \right){{\bf{C}}_{{n_{\rm{u}}}}\left(m,r\right)}} \right|}^2}} / {\left( {2{\sigma ^2}} \right)}}} \right)
  \end{array}, \label{eq11}
\end{equation}
where ${{\bf{\Psi }}_r}$ represents the possible codewords of all users sharing the $r$-th ORE, then $p_{{v_u} \to {f_r}}^{\left( {{k_{\rm{it}}} + 1} \right)}\left( {{{\bf{C}}_{{n_{\rm{u}}}}\left(m,r\right)}} \right)$ is updated to,
\begin{equation}
\begin{array}{l}
  p_{{v_u} \to {f_r}}^{\left( {{k_{\rm{it}}} + 1} \right)}\left( {{{\bf{C}}_{{n_{\rm{u}}}}\left(m,r\right)}} \right) = \gamma _{{v_u},r}^{\left( {{k_{\rm{it}}} + 1} \right)}\\
  \quad \quad \quad  \times \prod\limits_{j \in {\Omega _u}\backslash r} {p_{{f_r} \to {v_u}}^{\left( {{k_{\rm{it}}} + 1} \right)}\left( {{{\bf{C}}_{{n_{\rm{u}}}}\left(m,r\right)}} \right)} ,\forall m,\forall {n_{\rm{u}}},r \in {{\bf{\Omega }}_u}
  \end{array}, \label{eq12}
\end{equation}
where ${{\bf{\Omega }}_u}$ represents the ORE index corresponding to the $d_{\rm{v}}$ non-zero element positions of the codeword of the $n_{\rm{u}}$-th user, ${{\bf{\Omega }}_u}\backslash r$ represents ${{\bf{\Omega }}_u}$ except for the $r$-th ORE, and $\gamma _{{v_u},r}^{\left( {{k_{\rm{it}}} + 1} \right)}$ can be expressed as
\begin{equation}
\gamma _{{v_u},r}^{\left( {{k_{\rm{it}}} + 1} \right)}{\rm{ = }}{\left( {\sum\limits_{m = 1}^M {p_{{v_u} \to {f_r}}^{\left( {{k_{\rm{it}}}} \right)}\left( {{{\bf{C}}_{{n_{\rm{u}}}}\left(m,r\right)}} \right)} } \right)^{ - 1}}. \label{eq13}
\end{equation}

After $K_{\rm{it}}$ iterations, the estimated transmission codewords of the $n_{\rm{u}}$-th user can be expressed as
\begin{equation}
{{{\bf{\hat C}}}_{{n_{\rm{u}}}} ^{\left( {{k_{it}}} \right)}} = \arg \mathop {\max }\limits_{m = 1, \cdots M} \prod\limits_{j \in {\Omega _u}} {p_{{f_j} \to {v_u}}^{\left( {{k_{\rm{it}}}} \right)}\left( {{{\bf{C}}_{{n_{\rm{u}}}}\left(m,r\right)}} \right)} ,\forall n_{\rm{u}}. \label{eq14}
\end{equation}

The set of all user transmission codewords obtained by using the SCMA-IRS-MPA decoder is
\begin{equation}
{{\bf{\hat C}}_{{\rm{MPA}}}} = \left\{ {{{ {{{{\bf{\hat c}}}_1}} }^{\left( {{k_{\rm{it}}}} \right)}}, \cdots ,{{ {{{{\bf{\hat c}}}_{{N_{\rm{u}}}}}} }^{\left( {{k_{\rm{it}}}} \right)}}} \right\}. \label{eq15}
\end{equation}

The above is the MPA decoder of the SCMA-IRS scheme. We express the decoding computational complexity of the MPA decoder according to the number of addition operations and the number of multiplication operations. Therefore, the number of additions and multiplications required by the MPA detector are $R{d_{\rm{f}}}\left( {{M^{{d_{\rm{f}}}}}\left( {4{d_{\rm{f}}} + {K_{\rm{it}}} + 1} \right) + N - {K_{\rm{it}}}} \right) + 1$ and $R{d_{\rm{f}}}\left( {{M^{{d_{\rm{f}}}}}\left( {4{d_{\rm{f}}} + {K_{\rm{it}}}{d_{\rm{f}}} + 3} \right) + N + M{K_{\rm{it}}}\left( {{d_{\rm{v}}} - 1} \right)} \right) + {N_{\rm{u}}}M\left( {{d_{\rm{v}}} - 1} \right)$ respectively.

\section{Environment Sensing}
Contrary to the multiple access process, the algorithm proposed in this section can sense the environmental information with the data sent by the user has been decoded correctly. Since the distribution of scatterers in the environment is sparse, sensing environmental information is essential to solve the CS reconstruction problem. As shown in (\ref{eq5}), on the $r$-th ORE, the solution of environmental information can be expressed as
\begin{equation}
{\bf{\hat x}} = \arg \mathop {\min }\limits_{\bf{x}} {\left\| {\bf{x}} \right\|_1}\quad \quad {\rm{s}}{\rm{.t}}{\rm{.}}\quad {\left\| {{{\bf{y}}_r} - {{\bf{s}}_r}{{\bf{H}}_r}} \right\|_2} \le {\varepsilon _{\rm{x}}}, \label{eq16}
\end{equation}
where $\varepsilon _{\rm{x}}$ is the slack variable, ${{\bf{y}}_r} \in \mathbb{C}^{{N_{\rm{T}}} \times {N_{\rm{R}}}}$ is the symbol sequence received by the AP receiving antennas, $N_{\rm{T}}$ is the time sequence length, and ${{\bf{s}}_r} \in \mathbb{C} ^{{N_{\rm{T}}} \times {N_{\rm{u}}}}$ is the transmitted symbol sequence of $N_{\rm{u}}$ users. In the received signal model, when both the transmitted data ${\bf{s}}_r$ and the received data ${\bf{y}}_r$ are known, as shown in (\ref{eq6}), the channel coefficients ${{\bf{H}}\left({{n_{\rm{u}}},{n_R}}\right)}$ can be obtained by simply solving the linear equations. After performing the same analysis on all the receiving antennas of the AP, the channel coefficient ${\bf{H}}_r$ on the $r$-th ORE is solved.

As shown in (\ref{eq3}), the ${\bf{H}}_r^{{\rm{LOS}}}$ and ${\bf{H}}_r^{{\rm{IRS}}}$ in the channel coefficient ${\bf{H}}_r$ are composed of LOS channels. Meanwhile, the reflection characteristic matrix $\bf{\Theta} $ of the IRS used for assist communication is given, and only ${\bf{H}}_r^{\rm{s}}$ contains unknown environmental information. The $n_{\rm{u}}$-th row of ${\bf{H}}_r^{\rm{s}}$ is expressed as
\begin{equation}
{\bf{H}}_r^{\rm{s}}\left(n_{\rm{u}}\right) = {{\bf{x}}^{\rm{T}}}{\rm{diag}}\left( {{\bf{H}}_r^{{\rm{s3}}}} \left(n_{\rm{u}}\right) \right){\bf{H}}_r^{{\rm{s2}}}{\bf{\Theta H}}_r^{{\rm{s1}}}, \label{eq17}
\end{equation}
\begin{equation}
{\left( {{\bf{H}}_r^{\rm{s}}}\left(n_{\rm{u}}\right)\right)^{\rm{T}}} = {{\bf{A}}_r^{\rm{s}}}\left(n_{\rm{u}}\right){\bf{x}}, \label{eq18}
\end{equation}
where ${{\bf{A}}_r^{\rm{s}}}\left(n_{\rm{u}}\right) \in \mathbb{C}^{{N_{\rm{R}}} \times {N_{\rm{s}}}}$ is the known channel coefficient, which is also called the measurement matrix in the CS problem. For $N_{\rm{u}}$ users, the (\ref{eq18}) is expressed as the matrix form of the CS problem,
\begin{equation}
{\left[ {\begin{array}{*{20}{c}}
  {{{\left( {{\bf{H}}_r^{\rm{s}}}\left(1\right) \right)}^{\rm{T}}}}\\
  {{{\left( {{\bf{H}}_r^{\rm{s}}}\left(2\right) \right)}^{\rm{T}}}}\\
   \vdots \\
  {{{\left( {{\bf{H}}_r^{\rm{s}}}\left(N_{\rm{u}}\right) \right)}^{\rm{T}}}}
  \end{array}} \right]_{{N_{\rm{u}}}{N_{\rm{R}}} \times 1}} = {\left[ {\begin{array}{*{20}{c}}
  {{{\bf{A}}_r^{\rm{s}}}}\left(1\right)\\
  {{{\bf{A}}_r^{\rm{s}}}}\left(2\right)\\
   \vdots \\
  {{{\bf{A}}_r^{\rm{s}}}}\left(N_{\rm{u}}\right)
  \end{array}} \right]_{{N_{\rm{u}}}{N_{\rm{R}}} \times {N_{\rm{s}}}}}{\left[ {\bf{x}} \right]_{{N_{\rm{s}}} \times 1}}
 \end{equation} 
\begin{equation}
\Rightarrow  {{\bf{\tilde{H}}}_r^{\rm{s}}}  = {\bf{\tilde{A}}}_r^{\rm{s}}{\bf{x}}. \label{eq19}
\end{equation}

\subsection{Generalized Approximate Message Passing}
The GAMP Algorithm\cite{Rangan} solves the problem of CS sparse reconstruction by iterative decomposition. The above problem formula (\ref{eq19}) is abbreviated as ${\bf{y}} = {\bf{\Phi x}} + {\bf{w}}$, where ${\bf{\Phi }} \in \mathbb{C} ^{{M_\phi } \times {N_\phi }}$ is the CS measurement matrix and ${\bf{w}} \sim {\cal C}{\cal N}\left( {0,{\sigma ^{\rm{w}}}} \right)$ represents noise. In this article, we assume that the distribution of environmental scatterers information as a Bernoulli-Gaussian distribution in a limited interval which probability density function is expressed as
\begin{equation}
\begin{array}{l}
  {p_{X{\rm{|}}{\bf{Q}}}}\left( {x|{\bf{q}}} \right) = \left( {1 - \lambda  + \alpha } \right)\delta \left( x \right)\\
  \quad \quad \quad  + \lambda {\cal N}\left( {x|\theta ,{\sigma ^{\rm{x}}}} \right)\left[ {u\left( x \right) - u\left( {x - 1} \right)} \right]
  \end{array}, \label{eq20}
\end{equation}
where all parameters be expressed as ${\bf{q}} \buildrel \Delta \over = \left[ {\lambda ,\alpha ,\theta ,{\sigma ^{\rm{x}}}} \right]$, $\delta \left(  \cdot  \right)$ is the Dirac function, $\lambda $ is the sparsity coefficient, $\alpha {\rm{ = }}\int_{x \in \left( { - \infty ,0} \right] \cup \left[ {1, + \infty } \right)} {\lambda {\cal N}\left( {x|\theta ,{\sigma ^{\rm{x}}}} \right)} dx$. $\theta  \in \left[ {{\rm{0}},{\rm{1}}} \right]$ and ${\sigma ^{\rm{x}}}$ represent the mean and variance of the environmental scatterers information distribution respectively.

The GAMP algorithm has defined two parameterized functions ${g_{\rm{in}}}\left(  \cdot  \right)$ and ${g_{\rm{out}}}\left(  \cdot  \right)$ and the specific algorithm is shown in Algorithm 1. At this point, we will show how to specify the parameterized functions ${g_{\rm{in}}}\left(  \cdot  \right)$, ${g_{\rm{out}}}\left(  \cdot  \right)$, ${g'_{\rm{in}}}\left(  \cdot  \right)$ and ${g'_{\rm{out}}}\left(  \cdot  \right)$, based on the maximum posterior probability (MAP) estimation, the input function can be written as
\begin{equation}
{g_{\rm{in}}}\left( {\hat v,{\sigma ^{\rm{v}}},{\bf{q}}} \right) = \arg \mathop {\max }\limits_x {F_{\rm{in}}}\left( {x,\hat v,{\sigma ^{\rm{v}}},{\bf{q}}} \right), \label{eq22}
\end{equation}
\begin{equation}
{F_{\rm{in}}}\left( {x,\hat v,{\sigma ^{\rm{v}}},{\bf{q}}} \right) = \log {p_{X|{\bf{Q}}}}\left( {x|{\bf{q}}} \right) - \frac{1}{{2{\sigma ^{\rm{v}}}}}{\left( {\hat v - x} \right)^2}, \label{eq23}
\end{equation}
\begin{equation}
{g'_{\rm{in}}}\left( {\hat v,{\sigma ^{\rm{v}}},{\bf{q}}} \right) = \frac{\partial {g_{\rm{in}}}\left( {\hat v,{\sigma ^{\rm{v}}},{\bf{q}}} \right)}{\partial \hat v} = \frac{1}{{1 - {\sigma ^{\rm{v}}}\frac{\partial ^2}{\partial  x^2} {\rm log}\left[{p_{X|{\bf{Q}}}}\left( {x|{\bf{q}}} \right)\right]}}, \label{eq24}
\end{equation}
the output function can be expressed as
\begin{equation}
{g_{\rm{out}}}\left( {y,\hat p,{\sigma ^{\rm{z}}}} \right) = \frac{{y - \hat p}}{{{\sigma ^{\rm{w}}}} + {\sigma ^{\rm{z}}}}, \label{eq25}
\end{equation}
\begin{equation}
{g'_{\rm{out}}}\left( {y,\hat p,{\sigma ^{\rm{z}}}} \right) = \frac{\partial {g'_{\rm{out}}}\left( {y,\hat p,{\sigma ^{\rm{z}}}} \right)}{\partial y} =  - \frac{1}{{{\sigma ^{\rm{w}}}} + {\sigma ^{\rm{z}}}}. \label{eq26}
\end{equation}

\begin{algorithm}[htb]
  \caption{The GAMP Algorithm\cite{Rangan}}
  \begin{algorithmic}[1] 
  \REQUIRE
  Given measurement matrix ${\bf{\Phi }} \in {\mathbb{C}^{{M_\phi } \times {N_\phi }}}$ and sequence of measurement value ${\bf{y}} \in \mathbb{C} ^{{M_\phi } \times 1}$.
  \STATE
  \textbf{Initialization}: Set environment prior parameter $\bf{q}$. Defined ${g_{\rm{in}}}\left(  \cdot  \right)$ and ${g_{\rm{out}}}\left(  \cdot  \right)$ from (\ref{eq22}), (\ref{eq24}). Set $t_i = 0$, ${\bf{\hat s}}\left( { - 1} \right) = 0$, ${\hat x_{{n_\phi }}}\left( {{t_i}} \right) > 0$, $\sigma _{{n_\phi }}^{\rm{x}}\left( {{t_i}} \right) > 0$.

  \WHILE {$\sum\limits_{{m_\phi }} {\left( {{y_{{m_\phi }}} - {{\hat z}_{{m_\phi }}}\left( {{t_i}} \right)} \right)}  > \varepsilon_{\rm{t}} $, where $\varepsilon_{\rm{t}} $ is a given error tolerance value}
  \STATE
  For each $m_\phi $:

  $\sigma _{{m_\phi }}^{\rm{z}}\left( {{t_i}} \right) = \sum\limits_{{n_\phi }} {\Phi _{{m_\phi },{n_\phi }}^2} \sigma _{{n_\phi }}^{\rm{x}}\left( {{t_i}} \right),$

  ${\hat p_{m_\phi }}\left( {t_i} \right) = \sum\limits_{n_\phi } {\Phi _{{m_\phi },{n_\phi }}}{{\hat x}_{n_\phi }}\left( {t_i} \right) - \sigma_{m_\phi }^{\rm{z}}\left( t \right) {\hat s_{{m_\phi }}}\left( {{t_i} - 1} \right),$

  ${\hat z_{{m_\phi }}}\left( {{t_i}} \right){\rm{ = }}\sum\limits_{{n_\phi }} {{\Phi _{{m_\phi },{n_\phi }}}} {\hat x_{{n_\phi }}}\left( {{t_i}} \right).$
  \STATE
  For each $m_\phi $:

  ${\hat s_{{m_\phi }}}\left( {{t_i}} \right) = {g_{\rm{out}}}\left( {{t_i},{y_{{m_\phi }}},{{\hat p}_{{m_\phi }}}\left( {{t_i}} \right),\sigma _{{m_\phi }}^{\rm{z}}\left( {{t_i}} \right)} \right),$

  $\sigma _{{m_\phi }}^{\rm{s}}\left( {{t_i}} \right) =  - {g'_{\rm{out}}}\left( {{t_i},{y_{{m_\phi }}},{{\hat p}_{{m_\phi }}}\left( {{t_i}} \right),\sigma _{{m_\phi }}^{\rm{z}}\left( {{t_i}} \right)} \right).$
  \STATE
  For each $n_\phi $:

  $\sigma _{{n_\phi }}^{\rm{v}}\left( {{t_i}} \right) = {\left[ {\sum\limits_{{n_\phi }} {\Phi _{{m_\phi },{n_\phi }}^2\sigma _{{n_\phi }}^{\rm{s}}\left( {{t_i}} \right)} } \right]^{ - 1}},$

  ${\hat v_{{n_\phi }}}\left( {{t_i}} \right) = {\hat x_{{n_\phi }}}\left( {{t_i}} \right) + \sigma _{{n_\phi }}^{\rm{v}}\left( {{t_i}} \right)\sum\limits_{{m_\phi }} {{\Phi _{{m_\phi },{n_\phi }}}{{\hat s}_{{m_\phi }}}\left( {{t_i}} \right)}.$
  \STATE
  For each $n_\phi $:

  ${\hat x_{{n_\phi }}}\left( {{t_i}{\rm{ + 1}}} \right) = {g_{\rm{in}}}\left( {{t_i},{{\hat v}_{{n_\phi }}}\left( {{t_i}} \right),\sigma _{{n_\phi }}^{\rm{v}}\left( {{t_i}} \right),{\bf{q}}} \right),$

  $\sigma _{{n_\phi }}^{\rm{x}}\left( {{t_i}{\rm{ + 1}}} \right) = \sigma _{{n_\phi }}^{\rm{v}}\left( {{t_i}} \right){g'_{\rm{in}}}\left( {{t_i},{{\hat v}_{{n_\phi }}}\left( {{t_i}} \right),\sigma _{{n_\phi }}^r\left( {{t_i}} \right),{\bf{q}}} \right).$
  \STATE
  ${t_i} = {t_i} + 1.$
  \ENDWHILE
  \ENSURE
  Estimated sparse vector ${\hat x_{{n_\phi }}}\left( {{t_i}} \right)$ and $\sigma _{{n_\phi }}^{\rm{x}}\left( {{t_i}} \right)$.
  \end{algorithmic}
  \end{algorithm}

\subsection{Proposed Environment Sensing Algorithm}
In the process of solving the CS sparse reconstruction problem, the product $N_{\rm{u}}N_{\rm{R}}$ of the number of users and the number of receiving antennas, and the number of spatial pixels $N_{\rm{s}}$ are orders of magnitude different, that is ${N_{\rm{u}}}{N_{\rm{R}}} \ll {N_{\rm{s}}}$, so that the number of columns in the CS measurement matrix ${\bf{\tilde{A}}}_r^{\rm{s}}$ is much larger than the number of rows, and the compression ratio is too high. Therefore, the environmental information $\bf{x}$ cannot be recovered accurately. We improve the above algorithm to adapt to the continuous data stream sent from users in the proposed scenario. We use multiple data packets to recover the environmental information after multiple observations. The CS problem is redefined as
\begin{equation}
{\bf{H}}_r^{\rm{s}}\left(n_{\rm{u}},k\right) = {{\bf{x}}^{\rm{T}}}{\rm{diag}}\left( {{\bf{H}}_r^{{\rm{s3}}}}\left(n_{\rm{u}}\right) \right){\bf{H}}_r^{{\rm{s2}}}{{\bf{\Theta }}\left(k\right)}{\bf{H}}_r^{{\rm{s1}}}, \label{eq27}
\end{equation}
\begin{equation}
{\left( {{\bf{H}}_r^{\rm{s}}}\left(n_{\rm{u}},k\right) \right)^{\rm{T}}} = {\bf{A}}_r^{\rm{s}}\left(n_{\rm{u}},k\right){\bf{x}}, \label{eq28}
\end{equation}
\begin{equation}
{\left[ {\begin{array}{*{20}{c}}
  {{{\left( {{\bf{H}}_r^{\rm{s}}}\left(1,1\right) \right)}^{\rm{T}}}}\\
   \vdots \\
  {{{\left( {{\bf{H}}_r^{\rm{s}}}\left(n_{\rm{u}},k\right) \right)}^{\rm{T}}}}\\
   \vdots \\
  {{{\left( {{\bf{H}}_r^{\rm{s}}}\left(N_{\rm{u}},K\right) \right)}^{\rm{T}}}}
  \end{array}} \right]} = {\left[ {\begin{array}{*{20}{c}}
  {{\bf{A}}_r^{\rm{s}}}\left(1,1\right)\\
   \vdots \\
  {{\bf{A}}_r^{\rm{s}}}\left(n_{\rm{u}},k\right)\\
   \vdots \\
  {{\bf{A}}_r^{\rm{s}}}\left(N_{\rm{u}},K\right)
  \end{array}} \right]} \left[{\bf{x}}\right] , \label{eq29}
\end{equation}
\begin{equation}
\Rightarrow  \left[{{\bf{\tilde{H}}}_r^{\rm{s}}}\left(K\right)\right]_{{N_{\rm{u}}}{N_{\rm{R}}}K \times 1}  = \left[{\bf{\tilde{A}}}_r^{\rm{s}}\left(K\right)\right]_{{N_{\rm{u}}}{N_{\rm{R}}}K \times {N_{\rm{s}}}}\left[{\bf{x}}\right]_{{N_{\rm{s}}}\times 1}, \label{eq30}
\end{equation}
where ${\bf{\Theta }}\left(k\right)$ represents the IRS reflection characteristic matrix when the $k$-th data packet is received, and $K$ is the number of data packets. After multiple observations, the difference between the number of rows of the observation matrix $N_{\rm{u}}N_{\rm{R}}K$ and the number of columns $N_{\rm{s}}$ is relatively small, and environmental information can be sensed more accurately.

Since data packets are continuously transmitted during the communication process, and the amount of data is very large, it is impossible to store all the data packets $K$ sent at all times. We set a time sliding-window with a length of $n_{\rm{f}}$, store the received data packet $k$ at the current moment to the $n_{\rm{f}}$ data packets previously received, and use the data in the sliding-window for environment sensing.

Compared with the communication data that exists all the time, the data in the sliding-window is very limited. Therefore, a large amount of communication data transmitted earlier will be wasted in the process of environment sensing. To solve this problem, we propose a ``momentum-mode'', which combines the sensing results calculated at the previous moments to calculate the current environment sensing results and makes the current sensing result contain part of the information outside the sliding-window. According to (\ref{eq16}), (\ref{eq18}), the ``momentum-mode'' can be expressed as
\begin{equation}
{{{\bf{\hat x}}}_k}{\rm{ = }}\arg \mathop {\min }\limits_{\bf{x}} \left( {{{\left\| {\bf{x}} \right\|}_1}} \right){\rm{ + }}\mu {{{\bf{\hat x}}}_{k{\rm{ - 1}}}}, \label{eq31}
\end{equation}
\begin{equation}
{\rm{s}}{\rm{.t}}{\rm{.}}\quad {\left\|{ {{\bf{\tilde{H}}}_r^{\rm{s}}}\left(k\right) - {\bf{\tilde{A}}}_r^{\rm{s}}\left(k\right)}{\bf{x}} \right\|_2} \le {\varepsilon _{\rm{x}}}, \label{eq32}
\end{equation}
where $\mu $ is the momentum coefficient. The larger the momentum coefficient $\mu $, the more previous data information the current sensing result depends on. Therefore, the setting of the momentum coefficient $\mu $ should be set according to the actual system and will be further analyzed in section \uppercase\expandafter{\romannumeral7}.

\section{Joint Multi-User Detection and Environment Sensing Algorithm}
As mentioned in Section \uppercase\expandafter{\romannumeral3} and Section \uppercase\expandafter{\romannumeral4}, the accurate decoding of user communication data requires the knowledge of the environment, and if the data sent by the user is not recovered, the environmental information cannot be sensed accurately. Although a sufficient number of pilots can be used to implement the proposed environment sensing algorithm, an excessive number of pilots will cause a decrease in communication efficiency. To tackle this issue, we propose an iterative and incremental algorithm based on low-density pilots to jointly recover users' communication data and environmental information. Let ${{\bf{s}}_k}\left( {0 < k \le K} \right)$ denote the $k$-th data packets of all users. We insert a pilot $\bf{P}$ before each $K$ data packets, and the AP can obtain the received data ${\bf{y}}_{\rm{p}}$. According to the previous section, based on the pilot $\bf{P}$, the environmental information ${{\bf{\hat x}}_{\rm{p}}}$ can be roughly estimated. Meanwhile, we need to use subsequent communication data packets to further improve the environment sensing results. Therefore, we use the pilot $\bf{P}$, the received data ${\bf{y}}_{\rm{p}}$ and the estimated environmental information ${{\bf{\hat x}}_{\rm{p}}}$ as initial terms to start the iterative algorithm.

\subsection{The Proposed Iterative Algorithm}

\begin{figure*}
  \centering
  \includegraphics[width=6in]{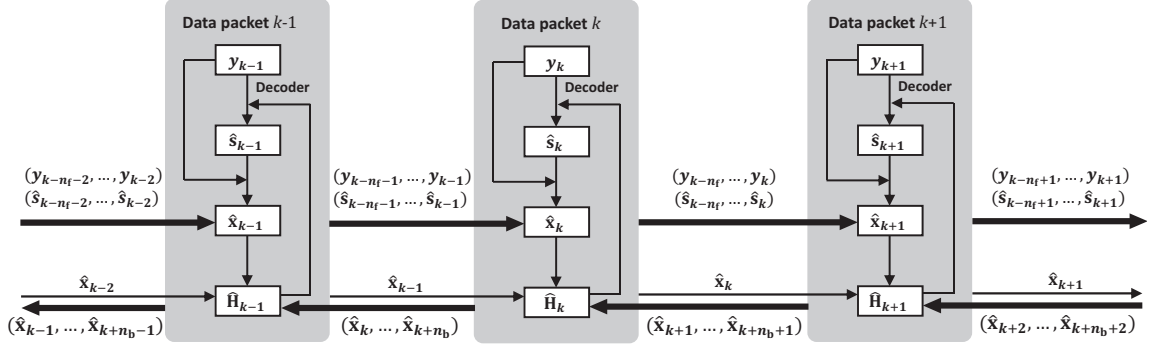}
  \caption{The proposed iterative algorithm.}
  \label{figIT}
  \end{figure*}

As shown in Fig. \ref{figIT}, after receiving the $k$-th packet data ${\bf{y}}_k$, the proposed iterative algorithm is divided into three parts:

1. Forward propagation: First, use the estimated environment information ${{\bf{\hat x}}_{k - 1}}$ after receiving the $(k-1)$-th data packet to estimate the current channel ${{\bf{\hat H}}_k}$, and then the decoder decodes the received data ${\bf{y}}_k$ to send data ${{\bf{\hat s}}_k}$ based on the estimation of current channel ${{\bf{\hat H}}_k}$. Finally, based on current ${\bf{y}}_k$, ${{\bf{\hat s}}_k}$, the received data ${{\bf{y}}_{k - {n_{\rm{f}}} - 1}}, \cdots ,{{\bf{y}}_{k - 1}}$ and the decoded data ${{\bf{\hat s}}_{k - {n_{\rm{f}}} - 1}}, \cdots ,{{\bf{\hat s}}_{k - 1}}$ in the previous $n_{\rm{f}}$ data packets, the current environmental information ${{\bf{\hat x}}_k}$ can be estimated more accurately. In the initial stage of the proposed algorithm, the results have not converged. Therefore, a certain number of iterations are required to make the system performance converge to a relatively accurate estimation of the transmitted data and environmental information, when $\left\lVert  {{{{\bf{\hat x}}}_k} - {{{\bf{\hat x}}}_{k - 1}}} \right\rVert _2 < {\varepsilon _{\rm{k}}}$, the ``momentum-mode'' is enabled.

2. Self-iteration: First, estimate the current environment ${{\bf{\hat x}}_k}$ by using the channel estimated ${{\bf{\hat H}}_k}$ in the forward propagation part. Then, the decoder decodes the currently transmitted data ${{\bf{\hat s}}_k}$ again and estimates the current environment information ${{\bf{\hat x}}_k}$ again. 
Finally, iterate $K_{\rm{s}}$ times to obtain more accurate transmission data and environmental information. $K_s$ can be set to a fixed value, or it can be gradually reduced as the algorithm converges, especially when $\left\lVert  {{{{\bf{\hat x}}}_k} - {{{\bf{\hat x}}}_{k - 1}}} \right\rVert_2 < {\varepsilon _{\rm{k}}}$, $K_s$ should be set to a small value. Enable the ``momentum-mode'' when $\left\lVert  {{{{\bf{\hat x}}}_k} - {{{\bf{\hat x}}}_{k - 1}}} \right\rVert_2 < {\varepsilon _{\rm{k}}}$.

3. Feedback: feedback the estimated environmental information ${{\bf{\hat x}}_k}$ in the $k$-th data packet to the previous $n_{\rm{b}}$ data packets, and based on the more accurate environmental information ${{\bf{\hat x}}_k}$ estimated by $k$-th data packet, estimate the received signal ${{\bf{\hat s}}_{k - {n_{\rm{b}}} - 1}}, \cdots ,{{\bf{\hat s}}_{k - 1}}$ again to improve the accuracy of decoding. Stop feedback when $\left\lVert {{{{\bf{\hat x}}}_k} - {{{\bf{\hat x}}}_{k - n_{\rm{b}} - 1}}} \right\rVert_2 < {\varepsilon _{\rm{k}}}$.

\begin{figure*}[htb]
  \centering
  \includegraphics[width=6in]{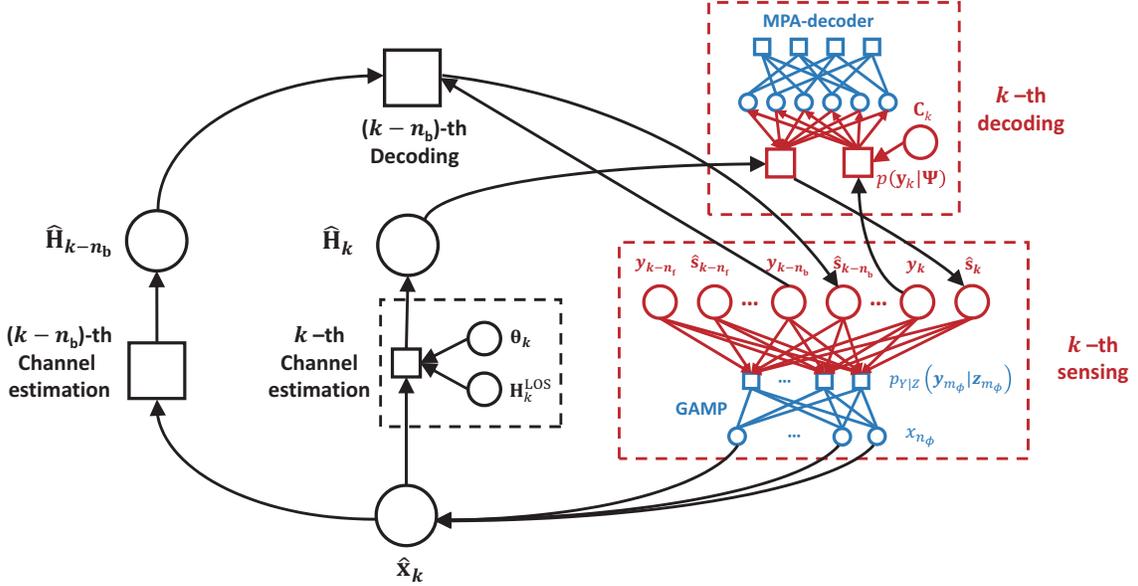}
  \caption{The Factor graph representation of the proposed iterative algorithm. }
  \label{figITGH}
  \end{figure*}

\begin{algorithm}[t]
  \caption{The proposed iterative algorithm}
  \label{alg2}
  \begin{algorithmic}[1] 
  \REQUIRE
  Calculated LOS channel matrix ${\bf{H}}_r^{{\rm{LOS}}}$, ${\bf{H}}_r^{{\rm{IRS1}}}$, ${\bf{H}}_r^{{\rm{s1}}}$, ${\bf{H}}_r^{{\rm{s2}}}$and ${\bf{H}}_r^{{\rm{s3}}}$. Given IRS reflection characteristic control matrix $\bf{\Theta}$.
  \STATE
  \textbf{Initialization}: Set pilot $\bf{P}$. Defined sliding-window length $n_{\rm{f}}$, $n_{\rm{b}}$. Set ${\varepsilon _{\rm{k}}} > 0$, $0 < \mu  < 1$, $K_{\rm{s}} > 0$.
  \STATE
  Estimate ${{\bf{\hat x}}_{\rm{p}}}$ from $\bf{P}$ and ${\bf{y}}_{\rm{p}}$ by GAMP. Let ${{\bf{\hat x}}_{\rm{0}}} = {{\bf{\hat x}}_{\rm{p}}}$, ${\bf{y}}_0 = {\bf{y}}_{\rm{p}}$, ${{\bf{\hat s}}_0} = {\bf{P}}$.
  \FOR {$k = 1, 2, \cdots ,K$}

  \STATE
  Estimate ${{\bf{\hat H}}_k}$ from ${{\bf{\hat x}}_{k - 1}}$ according to (\ref{eq3}).
  \STATE
  Estimate ${{\bf{\hat s}}_k}$ from ${\bf{y}}_k$ and ${{\bf{\hat H}}_k}$ based on SCMA-IRS-MPA decoder.
  \STATE
  Estimate ${{\bf{\hat x}}_k}$ from ${{\bf{y}}_{k - {n_{\rm{f}}} - 1}}, \cdots ,{{\bf{y}}_{k - 1}}$ and ${{\bf{\hat s}}_{k - {n_{\rm{f}}} - 1}}, \cdots ,{{\bf{\hat s}}_{k - 1}}$ according to formula (\ref{eq31}).
  \STATE
  Replace ${{\bf{\hat x}}_{k - 1}}$ with ${{\bf{\hat x}}_k}$, Repeat steps 3 to 5 $K_{\rm{s}}$ times.
  \STATE
  Estimate ${{\bf{\hat H}}_{k - 1}}, \cdots ,{{\bf{\hat H}}_{k - {n_{\rm{b}}} - 1}}$ from ${{\bf{\hat x}}_k}$ according to (\ref{eq3}).
  \STATE
  Estimate ${{\bf{\hat s}}_{k - 1}}, \cdots ,{{\bf{\hat s}}_{k - {n_{\rm{b}}} - 1}}$ from  ${{\bf{y}}_{k - 1}}, \cdots ,{{\bf{y}}_{k - {n_{\rm{b}}} - 1}}$ and ${{\bf{\hat H}}_{k - 1}}, \cdots ,{{\bf{\hat H}}_{k - {n_{\rm{b}}} - 1}}$ according to (\ref{eq31}).
  \STATE
  If $\left\lVert  {{{{\bf{\hat x}}}_k} - {{{\bf{\hat x}}}_{k - 1}}} \right\rVert_2 < {\varepsilon _{\rm{k}}}$, start ``momentum-mode'', else set $\mu  = 0$.
  \ENDFOR
  \ENSURE
  Estimated environment information ${{\bf{\hat x}}_k}$ and data packet ${{\bf{\hat s}}_1}, \cdots ,{{\bf{\hat s}}_K}$.
  \end{algorithmic}
  \end{algorithm}

We summarize the proposed iterative algorithm in Algorithm \ref{alg2}. During the execution process of the algorithm, the forward propagation process can be executed every time a new data packet is received. The self-iteration process can be executed as many times as necessary at any time, and the feedback process should be executed after the self-iteration process. Since the cached data cannot be too much, the forward propagation sliding-window size $n_{\rm{f}}$ and the feedback window size $n_{\rm{b}}$ need to be adjusted according to the actual system ability.

The effectiveness of the proposed algorithm can be explained by the message passing theory. Fig. \ref{figITGH} shows the factor graph of the proposed iterative algorithm. The decoding algorithm and the sensing algorithm use each other's solution results as side information to achieve their performance. As the number of iterations increases, the environmental information and data information contained in the received data are separated and recovered. The components of the proposed iterative algorithm also reflect this idea: forward propagation passes the previously sensed environmental information and received data information to the next time slot so that the algorithm can incrementally optimize performance based on the continuously received data. The self-iteration process is executed repeatedly and iteratively based on the existing data, and the environmental information in the received data is fully obtained. The feedback process passes more accurate environmental information to the previous time slot and reduces the error caused by inaccurate decoding and sensing in the initial stage of the proposed algorithm.

\subsection{Computational Complexity Analysis}
The computational complexity of the proposed algorithm is mainly composed of two parts: 
\begin{itemize}
    \item [(1)] 
    From the perspective of data decoding, we use the MPA decoder, whose computational complexity is $\mathcal{O} \left(Rd_{\rm{f}}M^{d_{\rm{f}}}+N_{\rm{u}}M\right)$. Compared with the ML decoder whose computational complexity is $\mathcal{O}\left(RC^{N_{\rm{u}}}\right)$, the computational complexity of the MPA decoder is lower. For example, when there are more users, the computational complexity of the MPA-based decoder just increases linearly. 
    \item [(2)] 
    From the perspective of environment sensing, we use the GAMP algorithm, whose computational complexity is $\mathcal{O}\left(N_{\rm{u}}N_{\rm{R}}KN_{\rm{s}}\right)$. Compared with the OMP algorithm, whose computational complexity is $\mathcal{O}(N_{\rm{u}}N_{\rm{R}}KN_{\rm{s}} + (\lambda N_{\rm{s}})^3 )$, it can be seen that the GAMP algorithm is a relatively low-complexity CS reconstruction algorithm.
\end{itemize}

In summary, during the execution of the algorithm, replace $K$ with the sliding window sizes $n_{\rm{f}}$ and $n_{\rm{b}}$.
The computational complexity of the proposed iterative algorithm is $\mathcal{O}( {R{d_{\rm{f}}}{M^{{d_{\rm{f}}}}}} $ $+ {N_{\rm{u}}}M + {N_{\rm{u}}}{N_{\rm{R}}}{n_{\rm{f}}}{n_{\rm{b}}}{N_{\rm{s}}} )$, where $n_{\rm{f}}$ and $n_{\rm{b}}$ can be controlled according to the convergence of the algorithm to save computing resources.
It can be seen that the computational complexity of the proposed algorithm is mainly determined by the number of users $N_{\rm{u}}$ and the SCMA codebook parameter $d_{\rm{f}}$. In contrast, if the OMP algorithm and the ML decoder are used to design the iterative algorithm, then its computational complexity will be $\mathcal{O}( R{C^{{N_{\rm{u}}}}} + {N_{\rm{u}}}{N_{\rm{R}}}{n_{\rm{f}}}{n_{\rm{b}}}{N_{\rm{s}}} + (\lambda N_{\rm{s}})^3  )$, which is much higher than our algorithm.

In addition, the low complexity of the proposed iterative algorithm is also reflected in the use of low-density pilots, which effectively reduces the time-frequency resources and computing resources consumed by the pilots.

\section{System Performance Analysis}
In this section, we analyze the influence of the number of users on the decoding results of communication data and the accuracy of environment sensing. After receiving the $k$-th data packet, we calculate the mean square error (MSE) between the estimated environmental information ${{\bf{\hat x}}_k}$ and the actual environmental information $\bf{x}$ to evaluate the accuracy of the environment sensing,
\begin{equation}
{\rm{MSE}} = \frac{1}{{{N_{\rm{s}}}}}\left\| {{{{\bf{\hat x}}}_k} - {\bf{x}}} \right\|_2^2, \label{eq33}
\end{equation}
where $N_{\rm{s}}$ is the total number of point clouds in the environment, and the SER between the decoding results ${{\bf{\hat s}}_k}$ and the original transmission data ${{\bf{s}}_k}$ is calculated to evaluate the accuracy of data decoding.

Based on the received data ${{\bf{y}}_k}$ and decoded data ${{\bf{\hat s}}_k}$, the essence of estimating the environmental information ${{\bf{\hat x}}_k}$ is to solve the CS sparse reconstruction problem, and (16) can be expressed as
\begin{equation}
{{{\bf{\hat x}}}_k} = \arg \mathop {\min }\limits_{\bf{x}} {\left\| {\bf{x}} \right\|_1}, \label{eq34}
\end{equation}
\begin{equation}
\begin{array}{l}
  {\rm{s}}{\rm{.t}}{\rm{.}}\quad {\left\| {{{\bf{y}}_{k-{n_{\rm{f}}},k}} - {{{\bf{\hat s}}}_{k-{n_{\rm{f}}},k}}{{\bf{x}}^{\rm{T}}} {{\bf{\tilde{A}}}^{\rm{s}}\left(\left[k-n_{\rm{f}},k\right]\right)} } \right\|_2}\\
  \quad  {\left\| { {\left( {{{\bf{s}}_{k-{n_{\rm{f}}},k}} - {{{\bf{\hat s}}}_{k-{n_{\rm{f}}},k}}} \right){{\bf{x}}^{\rm{T}}}{{\bf{\tilde{A}}}^{\rm{s}}\left(\left[k-n_{\rm{f}},k\right]\right)}} } \right\|_2} + {\varepsilon _{\rm{x}}}
  \end{array}, \label{eq35}
\end{equation}
where ${{\bf{y}}_{k-{n_{\rm{f}}},k}}$, ${{\bf{\hat s}}_{k-{n_{\rm{f}}},k}}$, and ${{\bf{\tilde{A}}}^{\rm{s}}\left(\left[k-n_{\rm{f}},k\right]\right)}$ represent the received packets, the decoding results, and the measurement matrix in the forward propagation window of size $n_{\rm{f}}$ respectively, ${\rm{SER}} \propto {\left\| { {\left( {{{\bf{s}}_{k-{n_{\rm{f}}},k}} - {{{\bf{\hat s}}}_{k-{n_{\rm{f}}},k}}} \right){{\bf{x}}^{\rm{T}}}{{\bf{\tilde{A}}}^{\rm{s}}\left(\left[k-n_{\rm{f}},k\right]\right)}} } \right\|_2}$. Therefore, when the decoding error rate (SER) increases, the constraint conditions of the sparse reconstruction problem become more slack, and the estimated environmental information error (MSE) increased. According to the theory of CS \cite{Donoho}, the theoretical upper bound of environment sensing accuracy is,
\begin{equation}
{\left\| {{\bf{x}} - {{{\bf{\hat x}}}_k}} \right\|_2} \le c \cdot {R_p} \cdot {\left( {\frac{{{N_{\rm{u}}} \cdot {N_{\rm{R}}} \cdot {n_{\rm{f}}}}}{{\log {N_{\rm{s}}}}}} \right)^{{1 \mathord{\left/
 {\vphantom {1 2}} \right.
 \kern-\nulldelimiterspace} 2} - {1 \mathord{\left/
 {\vphantom {1 p}} \right.
 \kern-\nulldelimiterspace} p}}}, \label{eq36}
\end{equation}
where $c > 0$ is a constant, ${\left\| {\bf{x}} \right\|_p} \le {R_p},\left( {0 < p < 2} \right)$ is the sparsity condition, ${\rm{MSE}} \propto {\left\| {{\bf{x}} - {{{\bf{\hat x}}}_k}} \right\|_2}$. Therefore, when the number of users $N_{\rm{u}}$ increases, the rank of the measurement matrix in the sparse reconstruction problem increases, and the environment sensing error (MSE) decreases. In addition, when the number of non-zero elements in the environment increases, $R_p$ increases, and the environment sensing error (MSE) increases.

When using MPA decoder to decode the received signal based on the ML theory, for a single ORE, let $C = N_{\rm{u}} \times M$ be the number of symbols in the codebook, then $N_{\rm{u}}$ users at each moment have ${C^{{N_{\rm{u}}}}}$ ways to send symbols. The theoretical upper bound of the average decoding error rate (SER) is,
\begin{equation}
{\rm{SER}} \le \frac{1}{{{C^{{N_{\rm{u}}}}}}}\sum\limits_{{{\bf{s}}_{\rm{a}}}} {\sum\limits_{{{\bf{s}}_{\rm{b}}},{{\bf{s}}_{\rm{a}}} \ne {{\bf{s}}_{\rm{b}}}} {P\left( {{{\bf{s}}_{\rm{a}}} \to {{\bf{s}}_{\rm{b}}}} \right)} }, \label{eq37}
\end{equation}
where $P\left( {{{\bf{s}}_{\rm{a}}} \to {{\bf{s}}_{\rm{b}}}} \right)$ represents the pairwise error probability (PEP) that the symbol ${\bf{s}}_{\rm{a}}$ is incorrectly decoded to ${\bf{s}}_{\rm{b}}$. In (\ref{eq37}), there are a total of $\left( {C - 1} \right){C^{2{N_{\rm{u}}} - 1}}$ items for summation. Therefore, when other conditions are the same, SER increases as the number of users $N_{\rm{u}}$ increases. Due to the random distribution of scatterers in the environment, the channels from the users to the AP can be expressed as Rayleigh fading channels. According to the ML theory, the decoding process in (\ref{eq8}) can be expressed as
\begin{equation}
  {{{\bf{\hat s}}}_k} = \arg \mathop {\min }\limits_{j \in {C^{{n_{\rm{u}}}}}}{\left\| {{{\bf{y}}_k} - {{{\bf{\hat H}}_k}{{\bf{s}}_k}\left( j \right)} } \right\|^2}, \label{eq38}
\end{equation}
\begin{equation}
  {\rm{s}}{\rm{.t}}{\rm{.}}\quad {{\bf{y}}_k} = {{{\bf{\hat H}}}_k}{{\bf{s}}_k} + \left( {{{\bf{H}}_k} - {{{\bf{\hat H}}}_k}} \right){{\bf{s}}_k} + {\bf{w}}, \label{eq39}
\end{equation}
where $\bf{w}$ represents Gaussian white noise with variance $N_0$. Due to the random distribution of scatterers in the environment, the interference caused by channel estimation errors is also Gaussian. Then the PEP in (\ref{eq37}) can be expressed as
\begin{equation}
P\left( {{{\bf{s}}_{\rm{a}}} \to {{\bf{s}}_{\rm{b}}}} \right) = {{\mathbb{E}}_{{{{\bf{\hat H}}}_k}}}\left[ {Q\left( {\sqrt {\frac{{{{\left\| {{{{\bf{\hat H}}}_k}\left( {{{\bf{s}}_{\rm{a}}} - {{\bf{s}}_{\rm{b}}}} \right)} \right\|}^2}}}{{{N_0} + \mathbb{D} \left( {\left( {{{\bf{H}}_k} - {{{\bf{\hat H}}}_k}} \right){{\bf{s}}_k}} \right)}}} } \right)} \right], \label{eq40}
\end{equation}
Where $Q$ is the error function. Therefore, the channel estimation error caused by inaccurate environmental information estimation will lead to an increase in the decoding symbol error rate (SER).
\begin{figure}
  \centering
  \includegraphics[width=7.5cm]{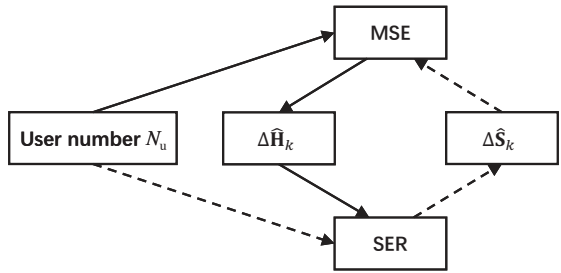}
  \caption{The trade-off relationship between the number of users and system performance.}
  \label{figtrade}
  \end{figure}

As shown in Fig. \ref{figtrade}, the solid line indicates that the increase in the number of users $N_{\rm{u}}$ promotes the accuracy of environment sensing, and the decrease of MSE reduces the channel prior information error $\Delta {{\bf{\hat H}}_k}$ for decoding, so the SER also decreases and the decoding becomes more accurate. On the other hand, the dotted line indicates that an increase in the number of users $N_{\rm{u}}$ increases the error of decoding (SER), and the decoded data error $\Delta {{\bf{\hat s}}_k}$ also increases, which increases the environment sensing error (MSE). Therefore, firstly, when the number of users decreases, the environment sensing error (MSE) increases. Secondly, when the number of users increases, the SCMA decoding error (SER) increases. Finally, because the proposed iterative algorithm repeatedly executes environment sensing and data decoding, their performance affects each other, resulting in the same trend of SER and MSE, the number of users should be a compromise choice. The number of users $N_{\rm{u}}$ has a trade-off relationship with system performance (MSE/SER) because the number of users $N_{\rm{u}}$ is traded for system performance. Finally, the optimal operating point could be estimated as
\begin{equation}
{\tilde N_{\rm{u}}} = \arg \mathop {\min }\limits_{{N_{\rm{u}}}} \quad {a_1} \cdot {\rm{MSE}} + {a_2} \cdot {\rm{SER}}. \label{eq41}
\end{equation}

When the number of users is ${\tilde N_{\rm{u}}}$, the system performance MSE and SER reach the best. In practical applications, we need to adjust the coefficients $a_1$ and $a_2$ to suit the system's requirements for communication and environment sensing performance. And select the appropriate SCMA codebook and system parameters according to the number of users in the scene to ensure that the number of actual users is within the optimal working range of the system.

\section{Numerical Results}
\begin{figure*}[t]
  \centering
  \subfigure[The original environment scatterer distribution.]{
  \includegraphics[width=0.4\textwidth]{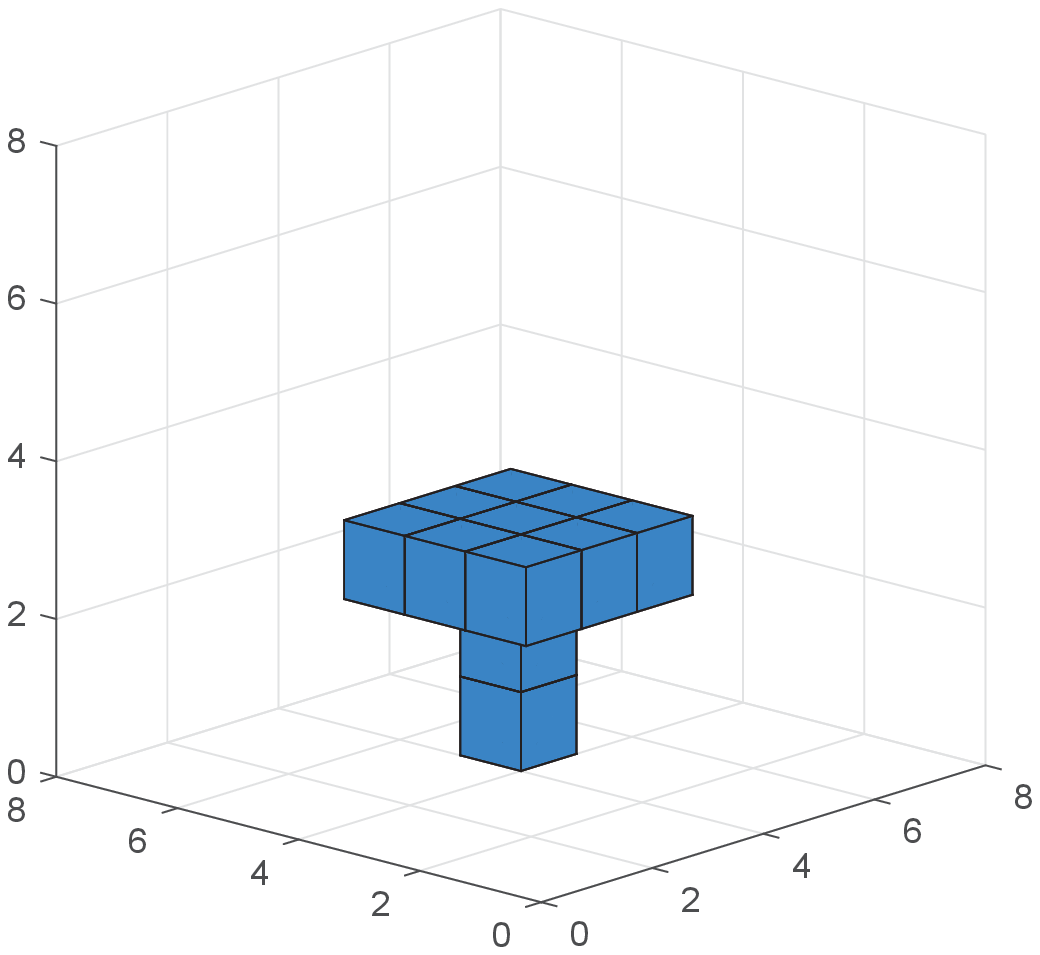}}
  \subfigure[The sensing result when $\rm{E_b/N_0}=0dB$.]{
  \includegraphics[width=0.4\textwidth]{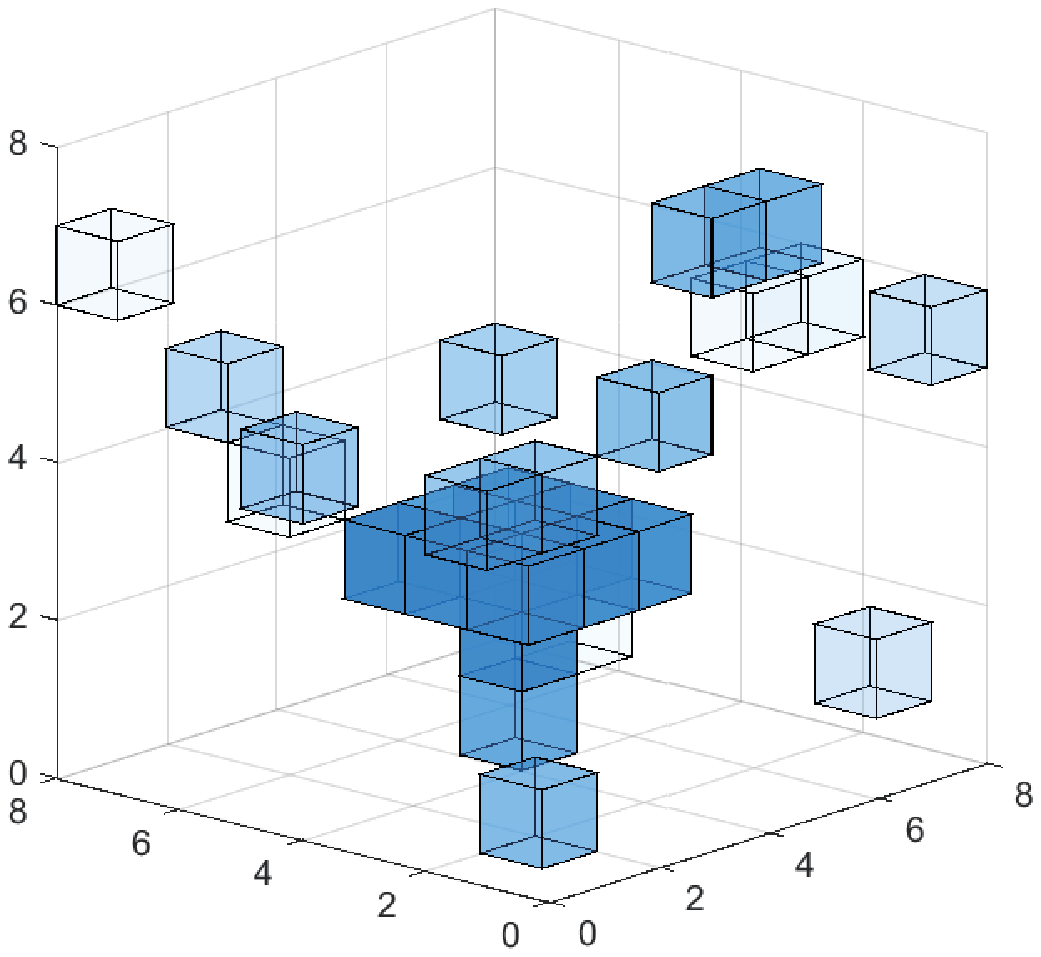}}
 \subfigure[The sensing result when $\rm{E_b/N_0}=5dB$.]{
  \includegraphics[width=0.4\textwidth]{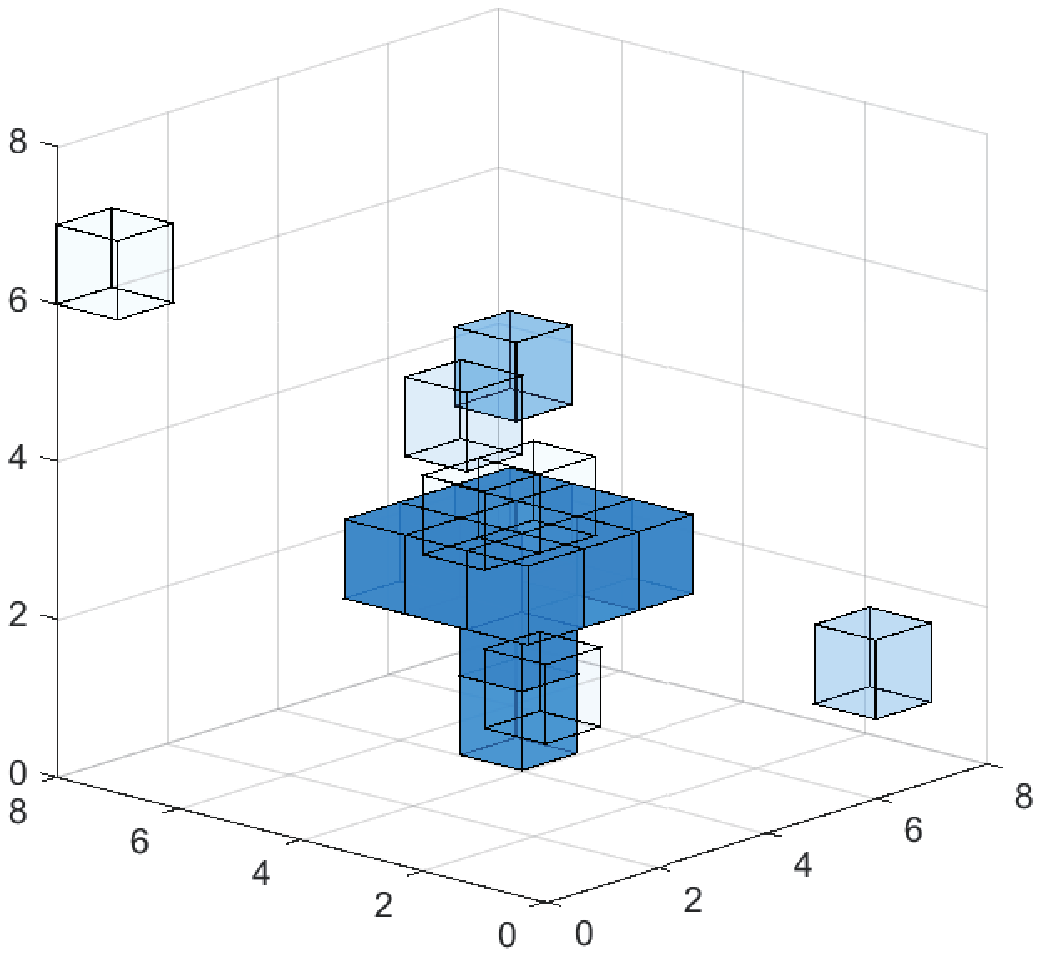}}
  \subfigure[The sensing result when $\rm{E_b/N_0}=10dB$.]{
  \includegraphics[width=0.4\textwidth]{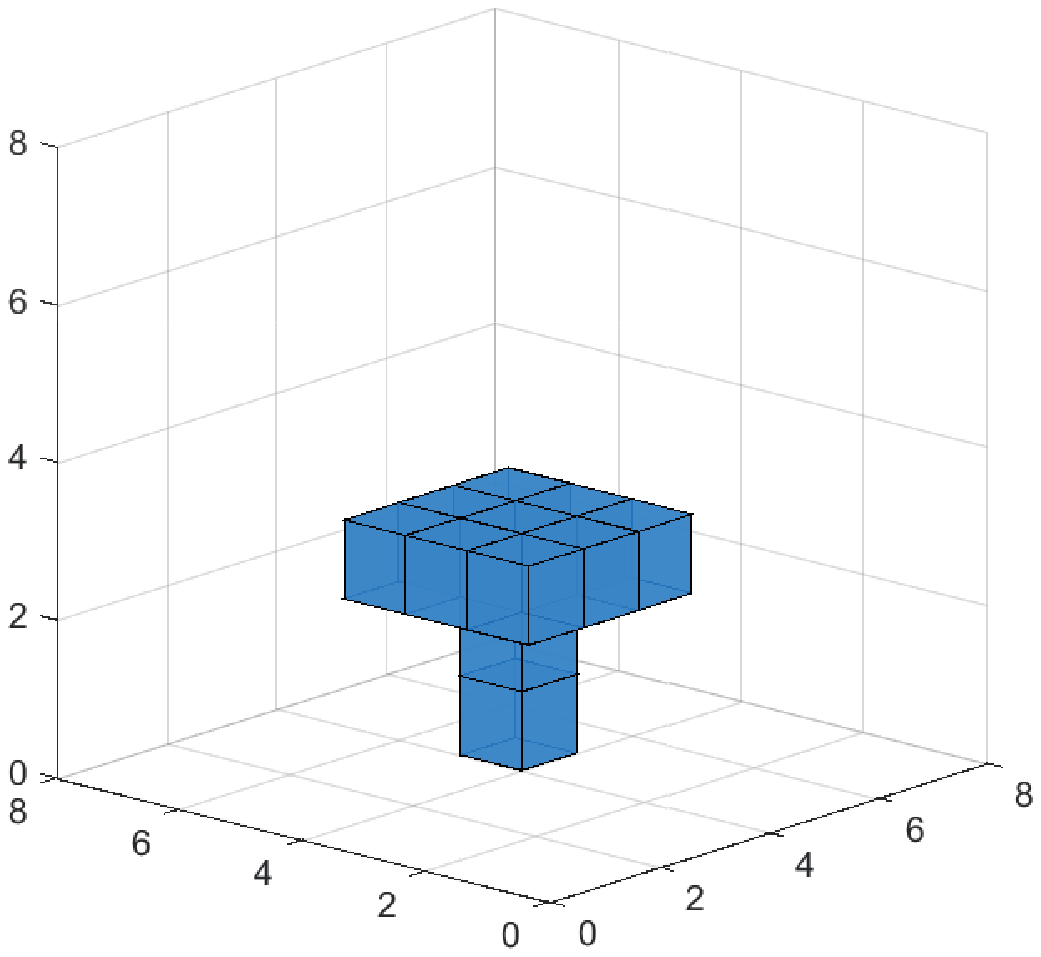}}
  \caption{The original environment scatterer distribution and the sensing results under different SNR conditions.}
  \label{figanswer}
  \end{figure*}

In this section, we simulated the performance of the algorithm and all simulations are conducted in MATLAB 2017b on a computing server with a Xeon E5-2697 v3 processer and 128GB memory. The simulation scenario is set in a room with a size of $4\rm{m} \times 4m \times 4m$, and the point cloud with a size of $8 \times 8 \times 8$ is used to represent the environmental information. The transmission signal frequency is set to 28 - 30GHz, and the bandwidth is 2GHz. A $20 \times 20$ IRS is used for assist communication. In order to meet the actual system design, we set the IRS amplitude reflection coefficient $\rho_{{n_{\rm{I}}}}  = 1$, the phase shift $\varphi_{{n_{\rm{I}}}}   = 0$ or $\pi $. The position of the scatterers distributed in the space is random, and the scattering coefficient ${\bf{x}}_{{n_{\rm{s}}}} \in \left[ {0,1} \right]$. As shown in Fig. \ref{figanswer}, small cubes are used to indicate the position distribution and scattering coefficient of the point cloud in space. The lower the transparency of the small cube, the larger the scattering coefficient of the point.

The distribution of the environment scatterer is shown in Fig. \ref{figanswer}(a), and the system parameters are set to the number of users $N_{\rm{u}}$ = 6, the number of OREs $R$ = 4, and $d_{\rm{v}}$ = 2. According to the convergence of the algorithm, we set $K_{\rm{s}} = 5$ in this section. After the proposed algorithm is iterated to convergence, the intuitive sensing results are shown in Fig. \ref{figanswer}(b), Fig. \ref{figanswer}(c), and Fig. \ref{figanswer}(d), when the signal-to-noise ratio (SNR, $\rm{E_b/N_0}$) are 0dB, 5dB, and 10dB respectively. It can be seen that the sensing result is very blurred when $\rm{E_b/N_0}=0dB$, and the shape of the target object can only be barely distinguished. As the SNR condition becomes better, the number of misidentified scatterers gradually decreases until $\rm{E_b/N_0}=10dB$ can clearly distinguish the shape of the target. This is because the SNR conditions affect the accuracy of communication data decoding, and therefore affect the accuracy of the sensing results.

\begin{figure}
  \centering
  \includegraphics[width=8cm]{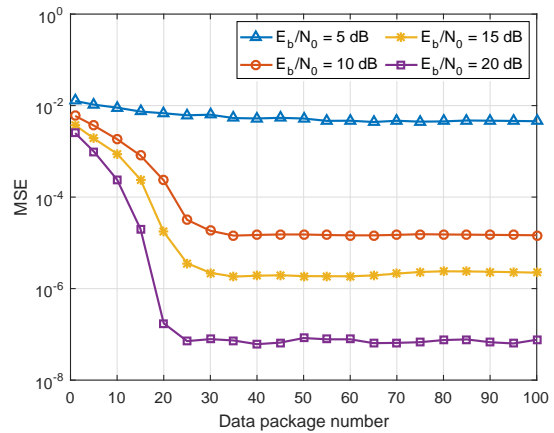}
  \caption{The relationship between the data packet number (number of iterations) and MSE.}
  \label{IT-MSE}
  \end{figure}

\begin{figure}
  \centering
  \includegraphics[width=8cm]{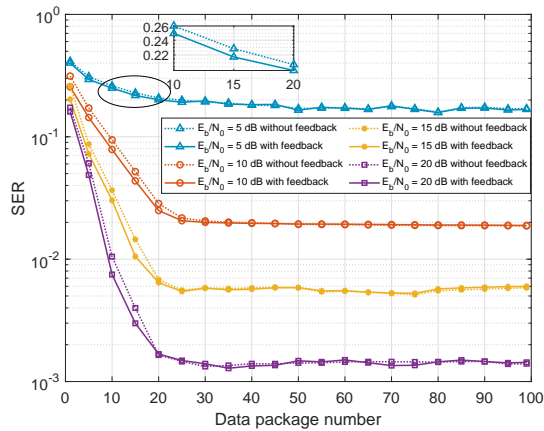}
  \caption{The relationship between the data packet number (number of iterations) and SER.}
  \label{IT-SER}
  \end{figure}

As shown in Fig. \ref{IT-MSE} and Fig. \ref{IT-SER}, the system parameters are set to the number of users $N_{\rm{u}}$ = 6, the number of OREs $R$ = 4, $d_{\rm{v}}$ = 2, and the sparsity of randomly generated environmental scatterers is 1.5\%. Set the forward propagation window size $n_{\rm{f}}$ = 10 and the feedback window size $n_{\rm{b}}$ = 1. We use MSE to evaluate the environmental sensing accuracy, when the number of data packets increases, the iterative algorithm converged, and the environment sensing result gradually becomes accurate. We use SER to evaluate the accuracy of transmission data decoding. As the number of data packets increases, the iterative algorithm converged, and the transmission data decoding results become more accurate. 
At the same time, due to the existence of feedback, as the number of data packets increases, the data packets at the previous time are decoded based on the more accurate environmental information at the later time. The feedback process improves the decoding accuracy during convergence.

\begin{figure}
  \centering
  \includegraphics[width=8cm]{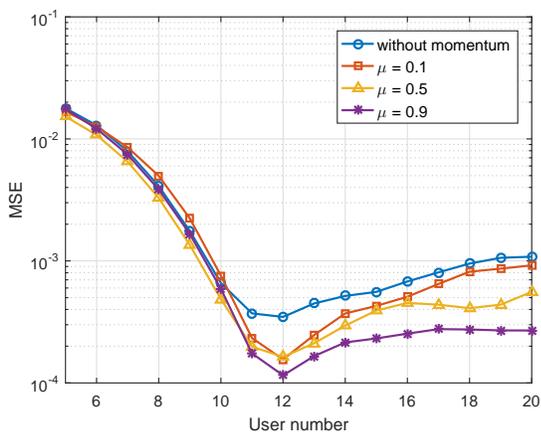}
  \caption{The trade-off relationship between the number of users and MSE.}
  \label{UE-MSE}
  \end{figure}

\begin{figure}
  \centering
  \includegraphics[width=8cm]{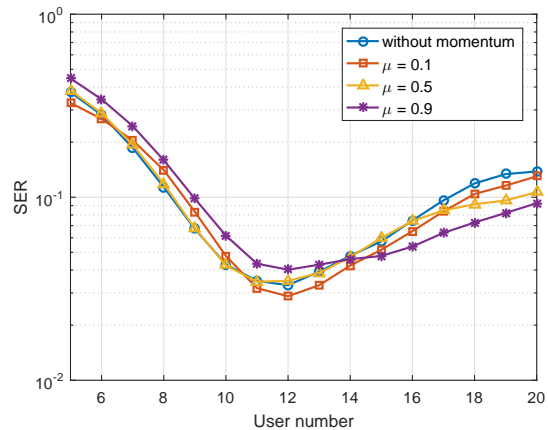}
  \caption{The trade-off relationship between the number of users and SER.}
  \label{UE-SER}
  \end{figure}

After the proposed algorithm iterates to convergence, the relationship between the number of users and system performance is shown in Fig. \ref{UE-MSE} and Fig. \ref{UE-SER}. We set a tough condition and the system parameters are set to the number of OREs $R$ = 7, $d_{\rm{v}}$ = 2, $\rm{E_b/N_0}=10dB$, and the sparsity of randomly generated environmental scatterers is 3\%. 
As analyzed in the section \uppercase\expandafter{\romannumeral6}, there is a trade-off relationship between the number of users and the system performance indicators SER and MSE. 
As the number of users changes, the environment sensing performance and the multi-user communication performance cannot reach the best at the same time.
As shown in Fig. \ref{UE-MSE} and Fig. \ref{UE-SER}, 
when the number of users is small, as the number of users increases, the sensing accuracy is significantly improved (as analyzed in \eqref{eq36}), and therefore the decoding accuracy is improved, until the optimal operating point ${\tilde N_{\rm{u}}} = 12$ is reached under simulation conditions. When the sensing of the environment is accurate sufficiently, a further increase in the number of users causes a decrease in the accuracy of decoding (as analyzed in \eqref{eq37} and \eqref{eq40}), and therefore the accuracy of environment sensing also decreases slightly.

Fig. \ref{UE-MSE} and Fig. \ref{UE-SER} also show the impact of ``momentum-mode'' on system performance. The ``momentum-mode'' uses data information outside the sliding-window for environmental sensing. As analyzed in Section \uppercase\expandafter{\romannumeral6}, the environment sensing accuracy is low when the number of users is small, so the ``momentum-mode'' accumulates errors and cannot improve system performance. However, ``momentum-mode'' can improve system performance when the number of users is higher than the optimal operating point, Since. It can be seen from the simulation results that when $\mu $ = 0.9, the momentum coefficient is large, the MSE and SER are greatly improved when there is a large number of users, and there is a negative impact on the SER in the case of few users. When $\mu $ = 0.5, the momentum coefficient is moderate, the MSE and SER indicators are slightly improved when there is a large number of users, and there is no significant impact on the system performance in the case of few users. When $\mu $ = 0.1, the momentum coefficient is small, there is no significant impact on the system performance. We recommend using a larger momentum coefficient when there is a large number of users, and using a moderate or small momentum coefficient in the case of few users.

Finally, in Table \ref{tab1}, we provide the run time of using the $k$-th received data package for data decoding and environment sensing (steps 4 to 10 in Algorithm \ref{alg2}). The simulation parameter settings are the same as those in Fig. \ref{UE-MSE} and Fig. \ref{UE-SER} and the momentum mode is not enabled. As shown in Table \ref{tab1}, we verify that the computational complexity of the algorithm increases with the increase in the number of users.
\begin{table}[h]
  \centering
  \caption{The run time of proposed algorithm.}
  \begin{tabular}{|p{1in}|p{0.3in}|p{0.3in}|p{0.3in}|p{0.3in}|}
      \arrayrulecolor{black}
      \hline  
      \makecell[l]{Number of users $N_{\rm{u}}$} & \makecell[c]{5} & \makecell[c]{10} & \makecell[c]{15} & \makecell[c]{20}\\
      \hline 
      \makecell[l]{Run time (s)} & \makecell[c]{4.95} & \makecell[c]{5.11} & \makecell[c]{5.33} & \makecell[c]{5.64}\\
      \hline 
      \end{tabular}
      \label{tab1}
  \end{table}

\section{Conclusion}
In the diverse wireless communication application scenarios in the future, environment sensing is an important component of the wireless communication system. In the scenario of IRS-assisted indoor uplink communication, we design a multiple access method and an environment sensing method. The multiple access method is based on SCMA. With the assistance of the IRS, based on the sparse codebook of the transmitted signals, the SCMA-IRS-MPA decoder is used. The environment sensing algorithm is based on the CS theory, including time sliding-window and ``momentum-mode'' which keep on sensing the environment while continuously receiving the data stream sent by the user. In this paper, the proposed multiple access algorithm and the proposed environment sensing algorithm rely on each other. Therefore, we propose a novel iterative algorithm based on low-density pilots to jointly solve the multiple access and environment sensing problems and achieve the integration of environment sensing and communication. Finally, numerical simulation has verified the convergence and effectiveness of the iterative and incremental algorithm and analyzed the trade-off relationship between the number of users and system performance. We also give a system parameter configuration method. The sensing-communication integration ideas and algorithms proposed in this paper will provide references for the development of new wireless communication technologies in the future.

\end{document}